\documentclass[namedreferences]{solarphysics}

\usepackage[hyperref,optionalrh,showbiblabels]{spr-sola-addons} 
\usepackage{graphicx}        
\usepackage{color}           
\usepackage{amsmath}
\usepackage{appendix}

\newcommand{\gammabeta}{[\gamma(t_i), \beta(t_i)]}

\begin{document}
\begin{article}
\begin{opening}

\title{Trajectory Determination for Coronal Ejecta Observed by WISPR/Parker Solar Probe}

\author[addressref=aff1]{\inits{}\fnm{P. C.}~\lnm{Liewer}}
\author[addressref=aff2]{\inits{}\fnm{J.}~\lnm{Qiu}}
\author[addressref=aff1]{\inits{}\fnm{P.}~\lnm{Penteado}}
\author[addressref=aff1]{\inits{}\fnm{J. R.}~\lnm{Hall}}
\author[addressref=aff3]{\inits{}\fnm{A.}~\lnm{Vourlidas}}
\author[addressref=aff4]{\inits{}\fnm{R. A.}~\lnm{Howard}}

\address[id=aff1]{Jet Propulsion Laboratory, California Institute of Technology, Pasadena, CA, 91109, USA (email: Paulett.Liewer@jpl.nasa.gov)}
\address[id=aff2]{Department of Physics, Montana State University, Bozeman, MT, 59717, USA}
\address[id=aff3]{Johns Hopkins University Applied Physics Laboratory, Laurel, MD, 20723, USA}
\address[id=aff4]{Naval Research Laboratory, Washington, DC, 20375, USA}
\runningauthor{Liewer et al.}

\begin{abstract}

The {\it Wide-field Imager for Solar Probe} (WISPR) on {\it Parker Solar Probe} (PSP), observing in white light, has a fixed angular 
field of view, extending from 13.5$^{\circ}$ to 108$^{\circ}$ from the Sun and approximately 50$^{\circ}$  in the transverse directions. 
Because of the highly elliptical orbit of PSP, the physical extent of the imaged coronal region varies directly as the distance from 
the Sun, requiring new techniques for analysis of the motions of observed density features. Here, we present a technique for determining 
the 3D trajectory of CMEs and other coronal ejecta moving radially at a constant velocity by first tracking the motion in 
a sequence of images and then applying a curve-fitting procedure to determine the trajectory parameters (distance vs. time, velocity, 
longitude and latitude). To validate the technique, we have determined the trajectory of two CMEs observed by WISPR that 
were also observed by another white-light imager, either LASCO/C3 or STEREO-A/HI1. The second viewpoint was used to verify the trajectory results from this new 
technique and help determine its uncertainty.

© 2020. All rights reserved.

\end{abstract}
\end{opening}

\section{Introduction}

The {\it Wide-field Imager for Solar Probe} \citep[WISPR;][]{Vourlidas2016} on {\it Parker Solar Probe} \citep[PSP;][]{Fox2016} 
is returning images of the corona from its unprecedented vantage point inside the orbit of Mercury. Launched in 
August of 2018, PSP flies in a highly elliptical heliocentric orbit, using seven Venus flybys to progressively decrease the perihelion 
and orbit period. The primary mission consists of 24 orbits, lasting to 2025, with the first perihelion at 35 R$_{\odot}$ and the 
final three perihelia at 9.86 R$_{\odot}$ from Sun-center. WISPR is a white-light instrument similar to the heliospheric imagers of the {\it Sun 
Earth Coronal Connection and Heliospheric Investigation} \citep[SECCHI;][]{Howard2008} on board the {\it Solar Terrestrial Relations 
Observatory} (STEREO) spacecraft.

The WISPR instrument has two telescopes, WISPR-I (inner; covering approximately $13.5^{\circ} - 53.0^{\circ}$ elongation from Sun center) and WISPR-O (outer; 
covering approximately $50.5^{\circ} - 108.5^{\circ}$ elongation from Sun center), 
with a combined angular field-of-view (FOV) extending 95$^{\circ}$ radially and approximately 50$^{\circ}$ transverse, with 
the inner edge pointed a fixed 13.5$^{\circ}$ from Sun center. Because of the changing distance to the Sun throughout the 
highly elliptical orbit, the physical extent of the imaged coronal region changes dramatically with time 
\citep[See visualization in Figure 1 of][ Paper I hereafter]{Liewer2019}.

Because the physical size of the region imaged by WISPR changes during the orbit, the observed motion of a white-light feature in 
a sequence of images is a combination of the feature's intrinsic motion and the spacecraft motion (Paper I). In this paper, 
we present and validate a technique for determining the trajectory (radial distance vs. time, latitude, longitude, and velocity) 
of a feature that is moving radially from the Sun at a constant velocity and that can be tracked in a sequence of WISPR images.  
It builds on a technique widely used in the analysis of data from previous coronagraphs and heliospheric imagers 
based on ``J-maps",  elongation vs. time maps at a fixed position angle 
around the Sun, where position angle is measured counterclockwise from solar north.  The trajectory is determined using the 
shape of elongation vs. time curves that moving white-light features make in such J-maps. The J-map 
technique was introduced in \cite{Sheeley1999} for the analysis of 
SOHO/LASCO data and widely used in the analysis of the STEREO/SECCHI data
\citep{Sheeley2008, Rouillard2008}. The method was adapted for use in the analysis of images from the wide-field SECCHI heliospheric 
imagers HI1 and HI2, when the assumption of motion in the plane-of-sky  breaks down and a correction for propagation out of 
the solar equatorial plane must be included  \citep{Rouillard2009, Savani2012}. 
The effect of spacecraft motion during the passage of a feature across the wide field of view of the 
heliospheric imagers has also been analyzed \citep{Conlon2014}.

The technique presented in this paper extends the J-map technique to the case of WISPR by 
taking into account the motions of both the feature and observer and, to first order, 
correcting for the approximately 4$^{\circ}$ inclination of the PSP orbit plane to 
the solar equatorial plane. The technique involves first tracking the 
feature in a time sequence of images, then fitting the track to analytic equations that 
relate the feature position in an image to its coordinates in a heliocentric inertial frame. 
We will refer to this method as the tracking/fitting technique. Note that because of the spacecraft motion and wide field-of-view, the assumption that a feature moving radially remains at a fixed solar position angle is no longer valid (Paper I).

To demonstrate the technique,  two CMEs were tracked and their 3-D trajectories determined. 
The tracking/fitting technique was validated, and its uncertainty investigated, using observations 
of the same feature from another white-light imager, either LASCO/C3 or STEREO-A/HI1 which provided a second view of the feature
at nearly the same time. The 3-D trajectory determined by this technique was 
used to predict where the tracked feature should be in an image from the other imager at a second 
viewpoint, and the predicted location was compared with the feature's observed location.  
One CME was seen during PSP's second orbit on 2019 April 2-3, probably originating from the newly emerged AR12737, 
and was moving at a significant angle to the solar equatorial plane. It was also observed by the 
SECCHI telescopes on STEREO-A; this second viewpoint confirmed our trajectory determination within the uncertainties. 
The second CME whose trajectory we determined was seen on the first orbit, 2018 November 1-2 \citep{Howard2019}; its origin and 
evolution was analyzed in detail in \cite{Hess2020}. This CME was traveling approximately in the solar equatorial plane. It was 
also observed by SOHO/LASCO and this second viewpoint was used to validate the technique and investigate its uncertainties.

The organization of this paper is as follows. Section 2 presents the tracking and fitting technique, using the two flux ropes as examples. The transformation of the tracks from pixel coordinates to the PSP orbital  frame, needed for fitting, is also explained. Section 3 gives the details of the two events and relates the predictions to observations of the same events by another white-light imagers.
Section 4 contains and summary and discussion. 
\section{Tracking and Fitting Technique}
\subsection{Coordinate Systems and Equations of Particle Trajectory}
\begin{figure}    
   \centerline{\includegraphics[width=1.0\textwidth,clip=]{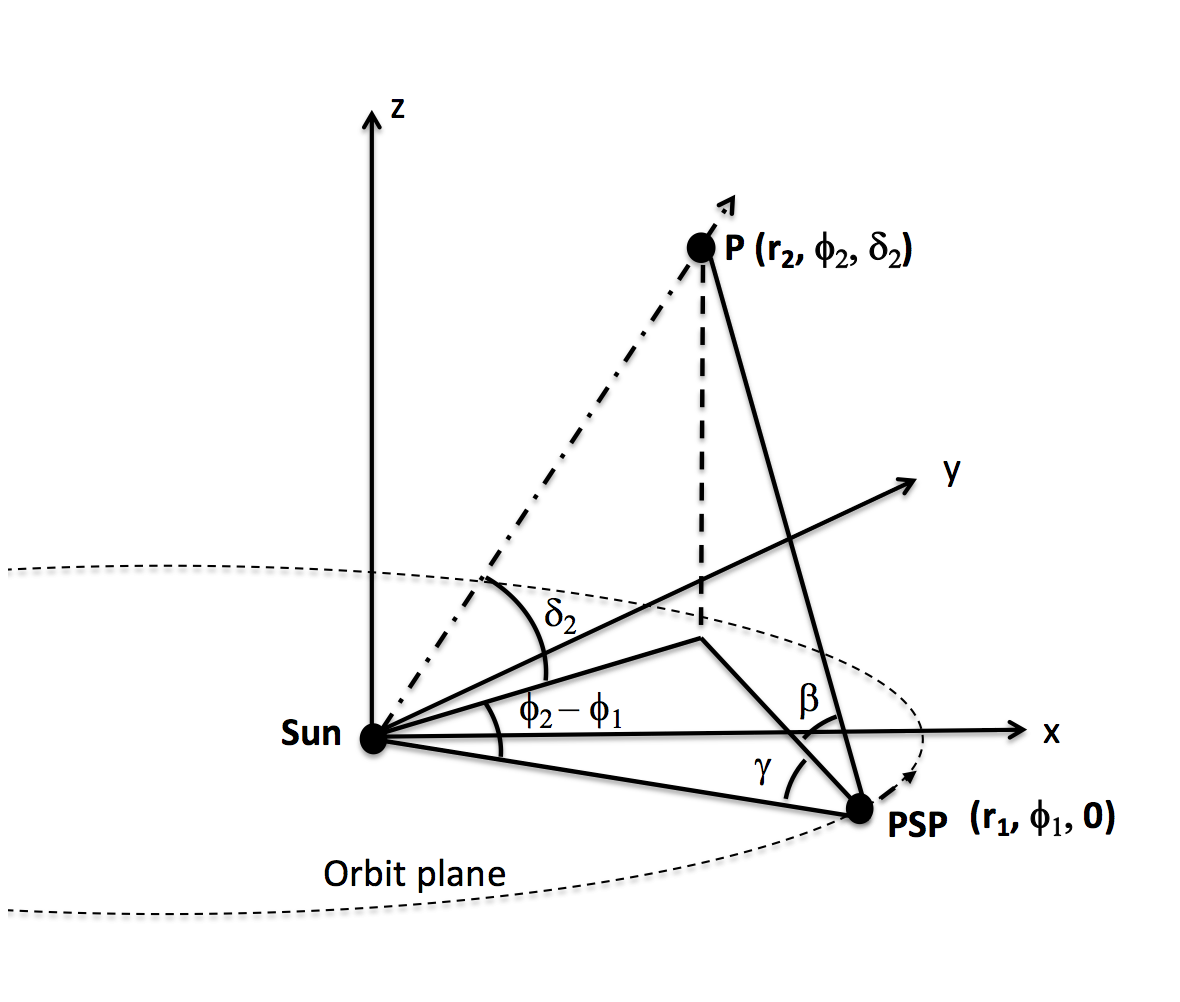}
              }
	\caption{Geometry relating the position of a feature P ($r2,\phi_2,\delta_2$) in the HCI coordinate frame to the two angles $\gamma,\beta$ defining the unique line-of-sight from PSP to the feature P under the assumption that PSP's orbit lies in the solar equatorial plane. (Adapted from Paper I.)}
   \label{fig:hci}
   \end{figure}

Paper I  introduced the framework that relates a feature's position in an WISPR image 
to its position in a heliocentric coordinate system under the assumption that the PSP orbit plane was 
in the solar equatorial plane. First, we defined a coordinate system referenced to the spacecraft position. 
Each pixel in the image defines a unique ray or line-of-sight from the spacecraft, specified by two angles, $\gamma$ and $\beta$, where 
$\gamma$ is the angle in the orbit plane, $\beta$ is the angle out of the orbit plane and Sun is at  ($\gamma, \beta) = (0,0)$. We then define the coordinates of
the heliocentric inertial (HCI) reference frame as $[r, \phi, \delta]$, where $r$ is the distance to the Sun, $\phi$ is the 
angle (longitude) in the solar equatorial plane (the $x-y$ plane), and $\delta$ is the angle 
(latitude) out of this plane. Figure~\ref{fig:hci}, based on Paper I (Figure~\ref{fig:11}), shows the 
relation of the line-of-sight from the spacecraft to a feature P, 
specified by $\gamma$ and $\beta$, to its location in the HCI frame. In this frame, 
the known (from the ephemeris) time-dependent coordinates of PSP (dashed line) are $[r_1, \phi_1, 0]$, and the coordinates 
of the feature P (dotted dashed line) are $[r_2, \phi_2, \delta_2]$. We assume that a feature of interest moves radially 
outward from the solar center and with a constant speed $V$, in which case its angles $\phi_2$ and $\delta_2$ are constant
in the HCI frame, and its radial coordinate can be written as $r_2(t) = r_2(t_0) + V*(t-t_0)$,  where $t_0$ is a reference time,
taken as the first time of the measurements. In the following text, we refer to $r_2(t_0)$ as $r_{20}$. Thus, the 
trajectory of the feature P is defined by four parameters: $r_{20}, V, \phi_2$, and $\delta_2$. Also shown 
in the figure is the separation angle between the feature P and PSP in the $x-y$ plane, $\phi_2-\phi_1$. 

Using basic trigonometry, we obtained two equations relating the coordinates introduced in Paper I.
These equations can be solved to determine the four trajectory parameters from a set of $\gamma(t)$ and $\beta(t)$ 
measurements of a feature's position in a sequence of WISPR images under the assumption that the PSP orbit plane lies in the solar equatorial plane.
\begin{flalign}
& \frac{\tan\beta (t)}{\sin\gamma (t)} = \frac{\tan\delta_2}{\sin[\phi_2 - \phi_1 (t)]}, \label{eq:1} \\
& \cot\gamma(t) = \frac{r_1(t) - r_2(t)\cos\delta_2 \cos[\phi_2 - \phi_1(t)]}{r_2(t)\cos\delta_2 \sin[\phi_2 - \phi_1(t)]}. \label{eq:2}
\end{flalign}
Equation ~\ref{eq:1} relates the feature's angles in the HCI frame to its angles $\gamma$ and $\beta$ referenced to the location of the PSP spacecraft; 
there is no dependence on 
either $r_1$ or $r_2$. For a feature moving in the solar equatorial plane ($\delta_2 = \beta=0$) and for a stationary spacecraft 
($r_1$ and $\phi_1$ constant in time), the second equation reduces to the starting equation used in J-map analyses to determine 
the trajectory of CMEs and other density features \citep{Sheeley1999, Sheeley2008, Rouillard2008, Conlon2014}.  

\begin{figure}    
\includegraphics[width=0.49\textwidth,clip=]{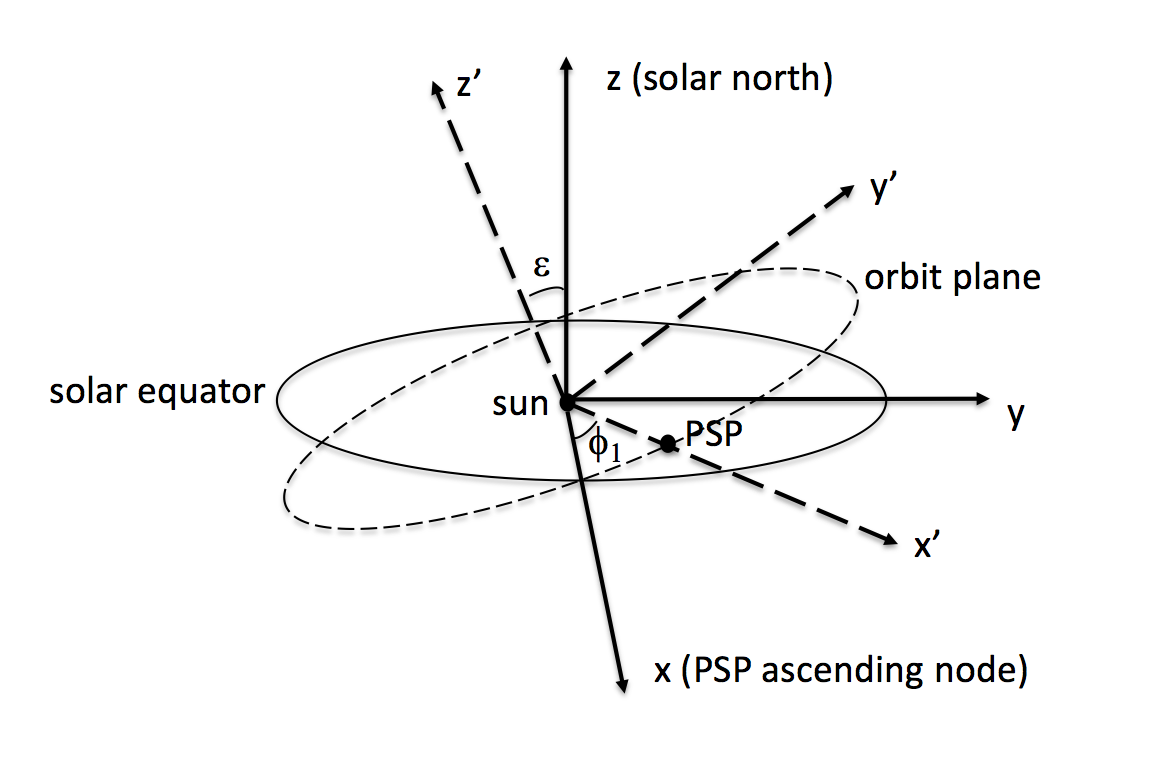}
   \includegraphics[width=0.49\textwidth,clip=]{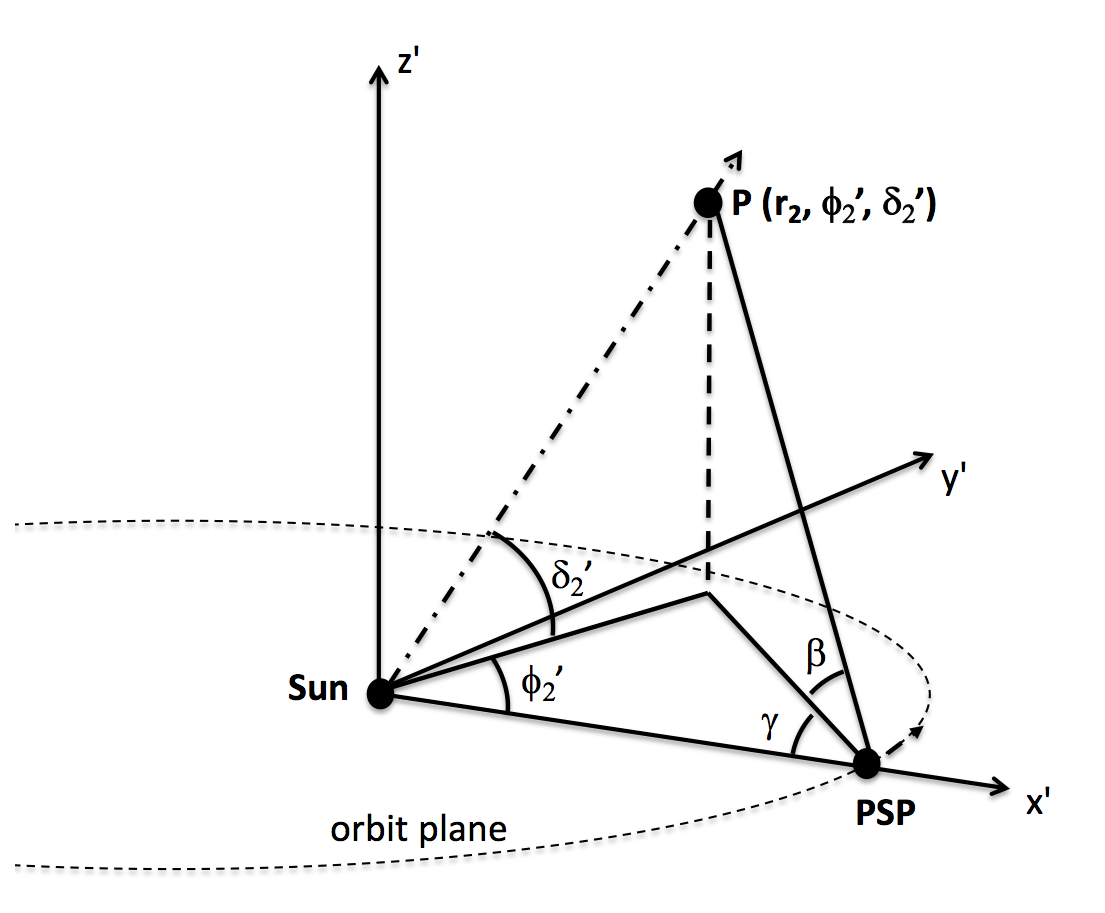}
   \caption{Left: Geometry showing the relation of the HCI frame to the PSP orbital  frame, where the inclination 
   $\epsilon$ between the solar equatorial and PSP orbit planes exaggerated for clarity. Solid lines show the HCI 
   coordinate  frame ($xyz$) in which the $z$-axis is solar north and the $x$-axis points to the ascending node of PSP's 
   orbit. Dashed lines show the heliocentic frame  ($x'y'z'$) where the $x'-y'$ plane 
is PSP's orbital plane and $x'$ points to PSP.   
Right: Geometry showing the coordinates of feature P in the  ($x'y'z'$) frame.}
   \label{fig:tran1}
   \end{figure}

Recall that these equations were derived assuming that the spacecraft orbit lies in the solar equatorial plane. This is a reasonable 
approximation since PSP's orbit is close to Venus' orbital plane, which is inclined by 3.8$^{\circ}$ to the solar equatorial plane. 
The inclination of PSP's orbit relative to the solar equatorial plane changes each time PSP uses Venus' gravity to reduce the 
perihelion. For the first two orbits, the period of time for the events analyzed in this paper, the inclination was about 4$^{\circ}$.
To take into account the inclination of the PSP orbit, we transform coordinates in the reference frame (referred as the $xyz$ frame),
in which the solar equatorial plane is the $x-y$ plane, to a heliocentric frame (the $x'y'z'$ frame), 
in which the $x'-y'$ plane is PSP's orbit plane and the $x'$ axis points to PSP's current location.
The left panel in Figure~\ref{fig:tran1} shows the relation between the two frames. For a feature located at $[r_2, \phi_2, \delta_2]$
in the $xyz$ frame, as illustrated in Figure~\ref{fig:hci}, its position in the $x'y'z'$ frame is illustrated in the
right panel of Figure~\ref{fig:tran1}.
Note that $\beta$  is the angle out of the PSP orbit plane and $\gamma$ is the angle in the plane. 
The angles are referenced to the position of the PSP spacecraft. 
We refer to this coordinate system as the PSP orbital frame.  The Sun is at ($\gamma, \beta) = (0,0)$, and thus  $\gamma$ reduces 
to a feature’s elongation (angle from the Sun) for motion in the PSP orbital plane.

We re-write the two equations relating $\gamma(t)$ and $\beta(t)$ to particle's trajectory in the $xyz$ frame (see Appendix~\ref{A1} for the full expressions). 
Since the orbit inclination $\epsilon$ is small, Equations~\ref{eq:1} and ~\ref{eq:2} can be modified to include $\sin\epsilon$ 
terms for first order corrections (see Appendix~\ref{A1}). 
\begin{flalign}
& \frac{\tan\beta(t)}{\sin\gamma(t)} = \frac{\tan\delta_2}{\sin[\phi_2 - \phi_1(t)]} \left(1 - F\sin\epsilon\right), \label{eq:3} \\
& \cot \gamma(t) = \frac{r_1(t)- r_2(t)\cos\delta_2\cos[\phi_2-\phi_1(t)]}{r_2(t)\cos\delta_2\sin[\phi_2 - \phi_1(t)]} 
\left(1 - G\sin\delta_2 \sin\epsilon\right).  \label{eq:4}
\end{flalign}
In these equations, $F$ is a function of the feature's angles $\phi_2$ and $\delta_2$, as well as PSP's 
angle $\phi_1(t)$, and $G$ is a function of the feature's position $r_2(t), \phi_2, \delta_2$ and PSP's position $\phi_1(t)$ and $r_1(t)$
(Equations~\ref{eq:F} and \ref{eq:G} in Appendix~\ref{A1}); $\epsilon$ is the inclination of the PSP orbit relative to the solar equatorial plane, and  $\phi_1(t)$ is PSP's angle in its
own orbit plane, measured from the ascending node of the PSP's orbit relative to the $x-y$ plane  (solar
equatorial plane, cf. Figure~\ref{fig:tran1}). Similarly, $\phi_2$ is the angle in the solar equatorial plane measured from this ascending node; therefore, $\phi_2$ here
is offset from the HCI longitude by a constant, which is the HCI longitude of the ascending node of PSP's orbit . 
Other parameters are defined the same way as in Equations~\ref{eq:1} and \ref{eq:2}. 

Employing these equations, we apply a least-squares curve-fitting algorithm to determine a feature's trajectory in the HCI frame from its positions
tracked in WISPR images, given a set of $\gammabeta$ measurements and their uncertainties at time $t_i$. At least two sets of $\gammabeta$ measurements are needed to 
determine the four unknown trajectory parameters. This technique includes the first-order corrections for the inclination of the PSP orbit relative to 
the solar equatorial plane, as described in Section 2.2.

\subsection{Curve Fitting Procedures}
To determine a features's trajectory in the HCI frame, we apply the procedure \verb!mpcurvefit.pro!,
availible from \verb!SolarSoft!, that performs Levenberg-Marquardt least-squares fit of WISPR measurements $\gammabeta$ at times $t_i$
to Equations~\ref{eq:3} and \ref{eq:4}. For out-of-plane motions, when the measured $\beta(t_i)$ is great than 4$^{\circ}$,
we perform a two-step fit. The first-step fit to Equation~\ref{eq:3} returns the two angles $\phi_2$ 
and $\delta_2$, and the second-step fit to Equation~\ref{eq:4} returns the initial distance $r_{20}$ 
and constant speed $V$. For in-plane motions, in which case the measured $\beta(t_i)$ approaches zero, we use 
a one-step fit to Equation~\ref{eq:4} to determine $r_{20}$, $V$, and $\phi_2$ altogether. Two examples are 
presented in the following text to illustrate the fitting techniques and results for out-of-plane motions and in-plane motions, respectively.

\subsubsection{Two-step Fit for Out-of-Plane Motions}
Equation~\ref{eq:3} relates a feature's position in the PSP orbit frame only to its angles
$\phi_2$ and $\delta_2$ in the $xyz$ frame, which does not depend on the feature's radial distance $r_2$. This allows
us to obtain the two angles $\phi_2$ and $\delta_2$ first, by the 
first-step fit of $\gammabeta$ measurements to Equation~\ref{eq:3}. Using
$\phi_2$ and $\delta_2$ determined from the first step, the second-step fit to Equation~\ref{eq:4}
returns $r_{20}$ and $V$. This approach reduces the number of free parameters in the non-linear fit, 
which is in general an ill posed problem. 

The convergence of a non-linear least squares fit often depends on the initial input.
We have developed a method to estimate the initial guess of $\phi_2$ to the zeroth 
order using Equation~\ref{eq:1}, without considering the first-order correction 
of the orbit inclination. The details of the method are given in Appendix~\ref{A2}. 
Applying this method to all the $\gammabeta$ measurements during the observation, 
we obtain the mean and deviation of the initial guess $\phi_2$, referred as $\bar\phi_2$ and $\Delta \phi_2$, 
respectively (see Appendix~\ref{A2}). For a given initial input of $\phi_2$ within the 
range of $\bar\phi_2 \pm \Delta \phi_2$,  
we also estimate $\delta_2, r_{20}$, and
$V$ to the zeroth-order using Equations~\ref{eq:1} and \ref{eq:2} (see Appendix~\ref{A2}).
These zeroth-order estimates of the four parameters are then used as initial input 
to fit $\gammabeta$ measurements, in two steps, to Equations~\ref{eq:3} and ~\ref{eq:4}, 
respectively, which take into account first-order corrections of orbit inclination, and return
more accurate determination of the parameters characterizing the feature's motion
in the HCI frame.

\begin{figure}    
   \centerline{\includegraphics[width=1.0\textwidth,clip=]{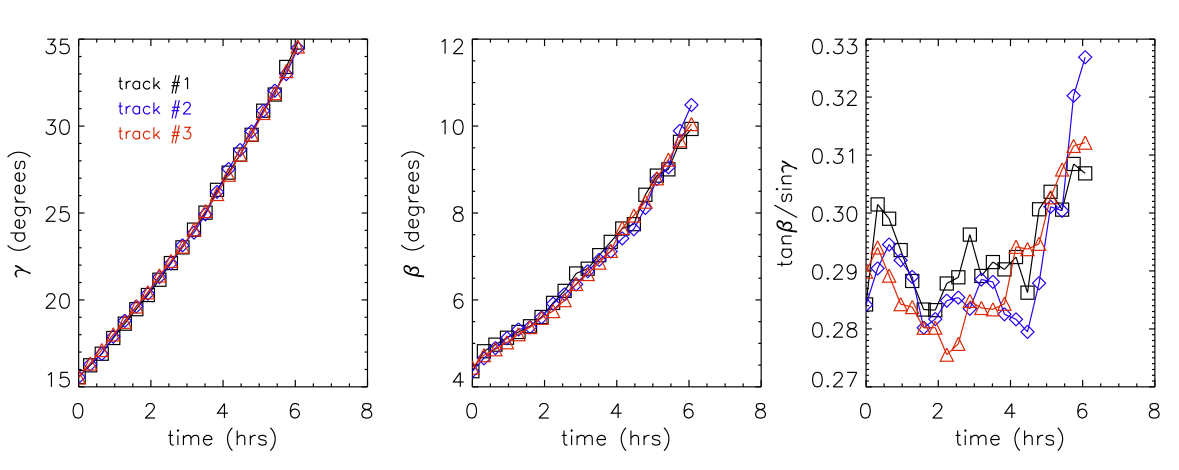}
              }
   \centerline{\includegraphics[width=1.0\textwidth,clip=]{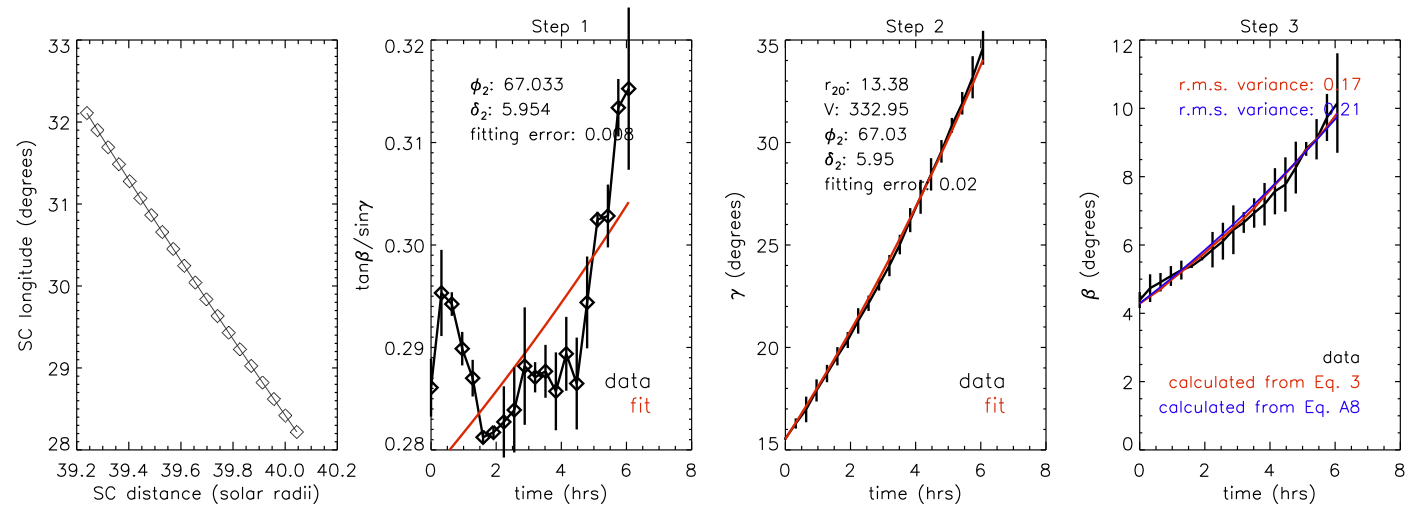}
              }
              \caption{Tracking and fitting results for the flux rope observed on 2019 April 2 (see Section 3.1 for details), which exhibits motion out of the orbit plane. 
The top row shows three independent trackings that return the $\gammabeta$ measurements. The bottom row shows the results of the two-step fit
using the mean and uncertainties of the $\gammabeta$ measurements. The left panel shows the PSP's longitude in the HCI frame
and distance to the Sun during the observation; the second panel shows the first-step fit to Equation~\ref{eq:3} to determine the two 
angles $\phi_2$ and $\delta_2$ of the tracked feature; the third panel shows the second-step fit to 
Equation~\ref{eq:4} to determine $r_{20}$ and $V$ of the feature; and the last panel compares the $\beta(t_i)$ calculated using Equation~\ref{eq:3} (red) 
or Equation~\ref{eq:a5} (blue) with observed values. Error bars (or uncertainties) in the plots are calculated using
$\gammabeta$ measurements from the three independent different, multiplied by five for clarity.  The quoted fitting error refers to the root-mean-square variance of the residuals of the fit.}
   \label{fig:ex1}
   \end{figure}

 Figure~\ref{fig:ex1} shows the $\gammabeta$ measurements in the top row and fitting results 
in the bottom row for a flux rope on 2019 April 2 (see Section 3.1 for the details of this event). 
The $\gammabeta$ measurements were obtained from three independent trackings, 
from which we derive the mean $\gammabeta$ and their deviations as measurement uncertainties
for each time $t_i$. 
We then conduct the two-step fit, weighted by measurement uncertainties, 
multiple times, each time varying the initial guess of $\phi_2$ in steps of 5$^{\circ}$ in the range $\bar\phi_2 - \Delta \phi_2$ 
to $\bar\phi_2 + \Delta \phi_2$, and the fitting parameters are determined from the fit with the smallest fitting errors.
For this data set, we calculated $\bar\phi_2  = 75^{\circ}$ and $\Delta \phi_2 = 43^{\circ}$ using the procedure described in 
the appendix.
In the bottom row of Figure~\ref{fig:ex1}, the second panel shows the result of the first-step fit 
to Equation~\ref{eq:3}, which returns $\phi_2$ and $\delta_2$ values (in HCI frame), and the 
third panel shows the result of the second-step fit to Equation~\ref{eq:4}, which returns $r_{20}$ and $V$ 
values. In each panel, the goodness of fit is indicated by the fitting error, or the root-mean-square variance of the residuals of the fit. 
The error bars represent the uncertainties in the measured $\gammabeta$, 
multiplied by five to make them more visible.  
The right-most panel plots the observed $\beta(t_i)$ in comparison with calculated values from 
Equation~\ref{eq:3} (red) and from Equation~\ref{eq:a5} (blue; see Appendix~\ref{A1}), as an 
independent check of the goodness of fit. For this event, from the best fit with the smallest fitting error, 
we have determined $\phi_2 = 67 \pm 1^{\circ}$ (in HCI frame), $\delta_2 = 6.0\pm 0.3^{\circ}$, 
$r_{20} = 13.38 \pm 0.01 R_{\odot}$, and $V = 333 \pm 1$ km s$^{-1}$, respectively. 
The uncertainty in each fitting parameter represents the 1-$\sigma$ error from the fitting procedure. The solution can be seen to be an excellent
fit to the input tracking data. Additional sources of error from
the tracking itself are discussed in Section 4.

\subsubsection{One-step Fitting for In-plane Motion}
The two-step fitting approach is applicable to motions out of PSP's orbit plane. If the measured
angle $\beta(t_i)$ is nearly zero, namely, the feature's motion is in PSP's orbit plane, the 
two-step fitting approach is no longer applicable. With the assumption of constant $\delta_2$,
Equation~\ref{eq:a1}, which relates the feature's motion in two frames, shows that
$\delta_2$ is also very small, $|\delta_2| \le |\epsilon|$. Therefore, the feature's motion is 
considered to be within the solar equatorial plane as well, with the maximum uncertainty of $\pm \epsilon$ in $\delta_2$.
In this case, $\beta$ will remain small $|\beta| \le \epsilon$ even when the spacecraft moves and its view of the feature
changes.  

For in-plane motion, to the zeroth order, $\delta_2 = 0$, and Equation~\ref{eq:3} becomes trivial and can no longer 
be used to determine $\phi_2$. We then determine the remaining three parameters $\phi_2$, $r_{20}$, and $V$
in a single fit of $\gamma(t_i)$ measurements to Equation~\ref{eq:4}, which degenerates to Equation~\ref{eq:2}  -- note
that, for $\beta(t_i), \delta_2 \sim 0$, the correction term in Equation~\ref{eq:4} becomes second-order. Therefore, no first-order
corrections are made in the determination of $\phi_2, r_{20}$, and $V$, when the feature's motion is nearly in the orbit plane.

For in-plane motion, we perform the one-step fit to Equation~\ref{eq:2} multiple times, 
each time using a fixed $\delta_2$ between $-4^{\circ}$ and 4$^{\circ}$ in 1$^{\circ}$ increments 
and a different initial input of $\phi_2$, to determine $\phi_2, r_{20}$, and $V$.  
The range of $\delta_2$ searched covers all possible values based on  Equation A.2.
It has been shown that the fit is sensitive to the initial 
input of $\phi_2$ but not sensitive to the initial input of $r_{20}$ and $V$. Therefore, 
for each $\delta_2$, we vary the initial input $\phi_2$ in a large range, but use the same initial 
input $r_{20} = 10$~R$_{\odot}$ and $v = 500$ km s$^{-1}$.

\begin{figure}    
   \centerline{\includegraphics[width=1.0\textwidth,clip=]{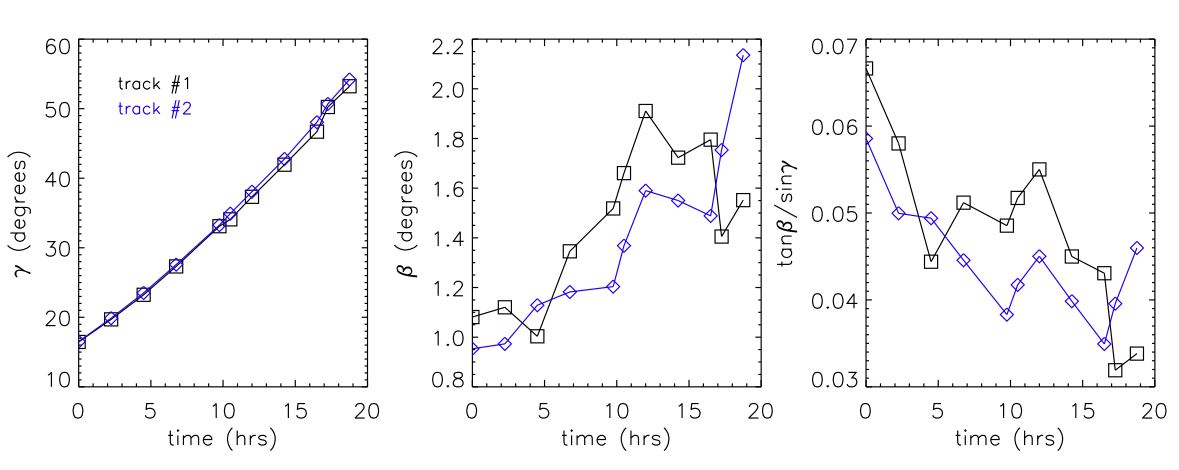}
              }
   \centerline{\includegraphics[width=1.0\textwidth,clip=]{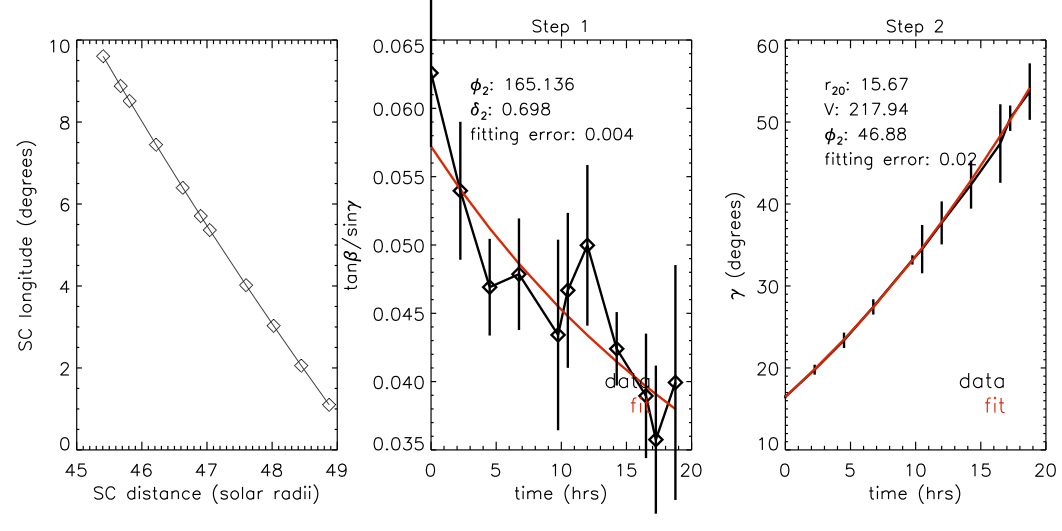}
              }
              \caption{Same as Figure~\ref{fig:ex1}; tracking (top) and fitting (bottom) results for the flux rope observed on 2018 November 1-2 
(see Section 3.2 for details), which exhibits motion nearly in the orbit plane. For this case, the first-step fit (the second panel
in the bottom row) returns a small value of $\delta_2$, thus the fitting results from the first-step are discarded, 
and $\phi_2, r_{20}$, and $V$ are determined from a one-step fit  to Eq.~\ref{eq:4} with a fixed $\delta_2$ value (the third panel in the bottom
row). $\delta_2$ is computed using the $\phi_2, r_0, V$ values from the one-step fit (see text in Section 2.2.2). For clarity, the error
bars plotted in the last panel are multiplied by five. } 
   \label{fig:ex2}
   \end{figure}

Figure~\ref{fig:ex2} shows the $\gammabeta$ measurements and the fitting results for a flux rope on 
2018 November 1 - 2 (see Section 3.2 for the details of this event). Shown in the top row, two 
independent trackings were obtained, yielding the mean and deviations of the $\gammabeta$ measurements. 
The bottom row shows the fitting results. Seen in the second panel, the first-step fit returns a 
small value of $\delta_2$; therefore, the fitting results from the first-step fit 
are deemed unreliable, and the returned $\phi_2$ value is discarded. Instead, we use a fixed $\delta_2$,
that increases by 1$^{\circ}$ from $-4^{\circ}$ to 4$^{\circ}$, and fit $\gamma(t_i)$ 
to Equation~\ref{eq:2} for three parameters, $r_{20}$, $V$, and $\phi_2$, as shown in the third panel. 
We note that the variation in $\delta_2$ results in negligible changes in the fitting results
of $\phi_2, r_{20},$ and $V$, and therefore the goodness of fit to Equation~\ref{eq:2} cannot be used to find the optimal
$\delta_2$. Instead, with $\phi_2$, $r_{20}$, and $V$ determined from the one-step fit, we compute $\beta(t_i)$ using 
Equations~\ref{eq:3} and also \ref{eq:a5}, to compare with observed $\beta(t_i)$ and determine $\delta_2$.

For the example shown in Figure~\ref{fig:ex2}, 90 fits have been conducted with $\delta_2$ varying
by 1$^{\circ}$ from $-4^{\circ}$ to 4$^{\circ}$, and 10 different initial guesses of $\phi_2$  in the range $10^{\circ} - 90^{\circ}$  for each $\delta_2$. 
The best fit to Equation~\ref{eq:2} yields the parameters $r_{20} = 15.7\pm 0.2 R_{\odot}$, $V = 218 \pm 2$ km s$^{-1}$, and
$\phi_2 = 47 \pm 2^{\circ}$ (in HCI frame). We then compute the mean and deviation of $\delta_2$ weighted by 
the root-mean-squares variance of the residuals in $\beta(t_i)$, yielding, $\delta_2  = -1 \pm 2^{\circ}$. It can be seen from Figure~\ref{fig:ex2} that the
solution is a good fit to the data.

\subsection{Tracking and Transformation from Pixel to PSP Orbital Frame Coordinates}

This section describes the method used to obtain a set of $\gammabeta$ measurements at times $t_i$,  
to be used as input to the curve fitting procedures described in Section 2.2. We use a time sequence of 
WISPR FITS images, binned to 960 by 1024 pixels, in the tracking software. 
These are running-differenced images created from Level 2 images, {\it e.g.,} the data has  been calibrated in units of Mean Solar Brightness (MSB), with bias, stray light and vignetting corrections applied, but no background subtracted. (The publicly released data, and the descriptions of the data products, can be found at \verb!https://wispr.nrl.navy.mil/wisprdata!.)
The tracking 
is done by manually placing the cursor on the feature being tracked in each image; the software reads and 
saves the pixel coordinates $[u(t_i),v(t_i)]$ at each time $t_i$, as well as the relevant information 
from the image's FITS header. Figure~\ref{fig:5} shows four of the WISPR-I images used for tracking the 
flux rope of  2019 April 2, with the location of the tracked feature marked by a red symbol (X). Here, we track the 
lower dark ``eye" of the skull-like flux rope. 
These images are in the inner camera frame of reference, not the PSP orbital frame. Thus the orbit plane is a curve across the image above the midplane, similar to the curve of  the orbit plane ($\beta =0 $) shown in the upper panel of Figure~\ref{fig:6}.

The pixel coordinates $[u(t_i), v(t_i)]$ are then transformed into the angular coordinates  $\gammabeta$ where the angles 
are measured in the PSP orbital  frame, that is, $\beta$ is the angle out of the PSP orbit plane and $\gamma$ is the angle 
in the orbit plane with the Sun at $[\gamma, \beta]= [0,0]$ as discussed in Section 2.1. The transformation 
takes into account the spacecraft's location and attitude as well as the camera projection and distortion effects. 
The necessary information is obtained from the image file's FITS header \citep[time of observation, camera model, 
World-Coordinate-System information; see ][]{Thompson2006}, used in conjunction with the various PSP SPICE  
kernels (orbit, attitude, etc.). Since the spacecraft is moving, the transformation is time dependent. The upper panel of Figure~\ref{fig:6}
is a full FOV WISPR image of the 2018 November 1-2 flux rope. Here, the simultaneous images from WISPR-I and WISPR-O 
have been projected to the WISPR-I camera frame (black). In this frame, the angles are referenced to the center 
of the inner telescope FOV and distortion and projection corrections have been made. 
Overlaid on this image is the grid of the $\gamma-\beta$ angles of the PSP orbital  frame (red). The lower panel of Figure~\ref{fig:6} shows 
the same image, now projected into the PSP orbital frame, with the PSP orbit plane defining $\beta = 0$. This 
projection illustrates the  transformation of pixels measurements $[u(t_i), v(t_i)]$ 
to angular coordinates $\gammabeta$ in the PSP orbital  frame for input to the fitting program described above.
Like the HelioProjective-Cartesian coordinate system, the PSP orbital frame is cartesian and  observer-centric. The PSP orbital frame system uses the Sun-spacecraft vector and the PSP velocity vector to define the orbit plane and the frame, whereas the HPC system uses the Sun-spacecraft vector and a vector perpendicular to this in the plane containing both the Sun-spacecraft vector and the solar North pole axis\citep{Thompson2006}. In both systems, the direction of the solar north vector varies.

We tested this tracking/fitting technique thoroughly for features moving both in and out of the plane orbit using sequences 
of synthetic WISPR images such as those described in Paper I and in \citet{Nistico2020}. 
For all cases, the trajectories used to create moving features in the synthetic images were returned accurately by the fitting program.

\section{Results and Validation}

This section describes the trajectory results obtained from application of the tracking/fitting technique to two CMEs observed
by WISPR, as well as the validation of the technique and investigation of its uncertainties using nearly simultaneous observation from a second 
spacecraft. 

\subsection{Flux rope of 2019 April 1-2}

\begin{figure}    
   \centerline{\includegraphics[width=1.0\textwidth,clip=]{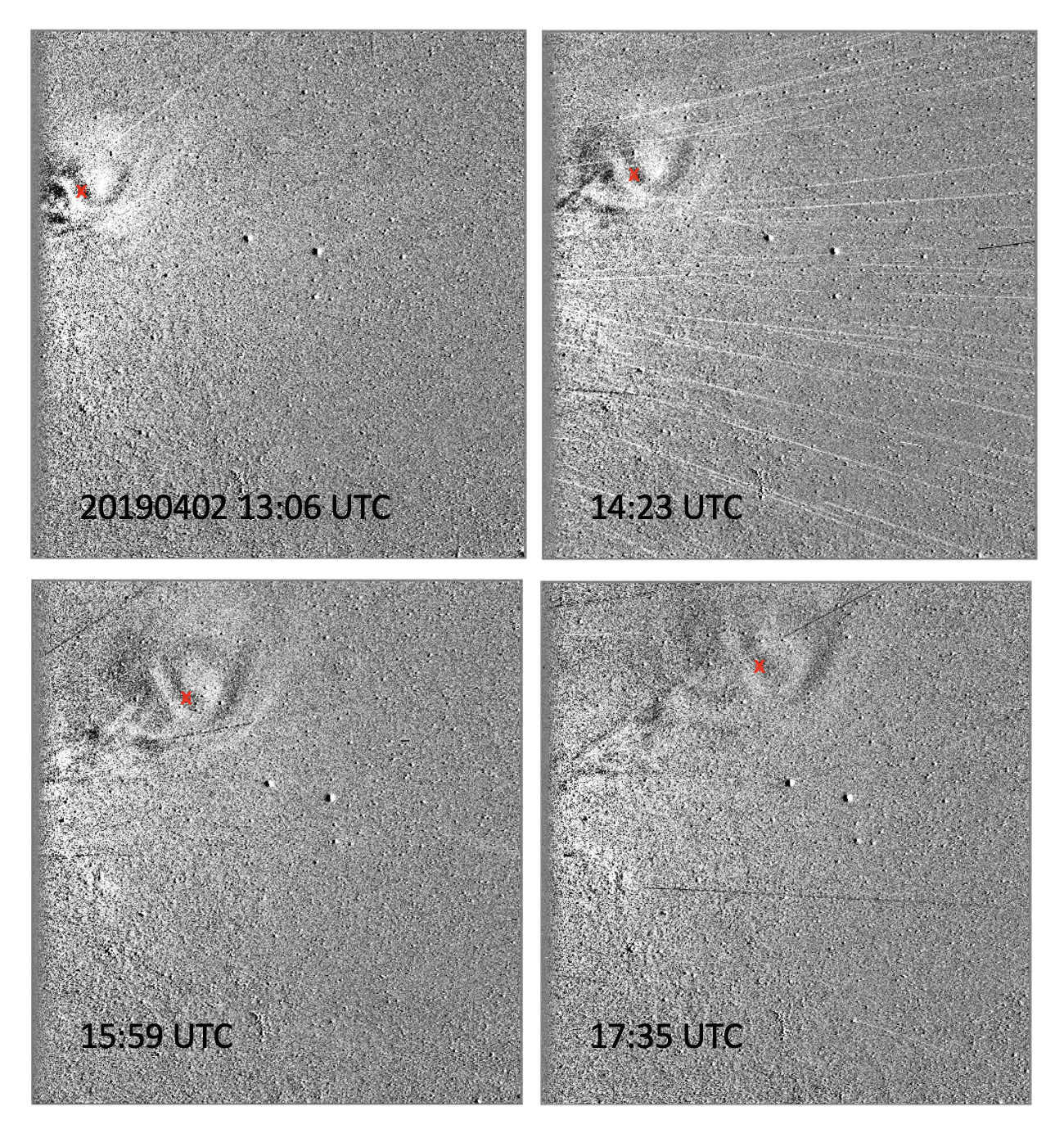}
 	      }
              \caption{WISPR-I running-difference images at four times for the CME of 2019 April 2. The
tracked feature, the lower dark ``eye" is marked with red X's. 
The streaks seen especially in the image at 14:23 are caused by dust created by impacts on the spacecraft.} 
   \label{fig:5}
   \end{figure}

\begin{figure}    
\includegraphics[width=1.0\textwidth,clip=]{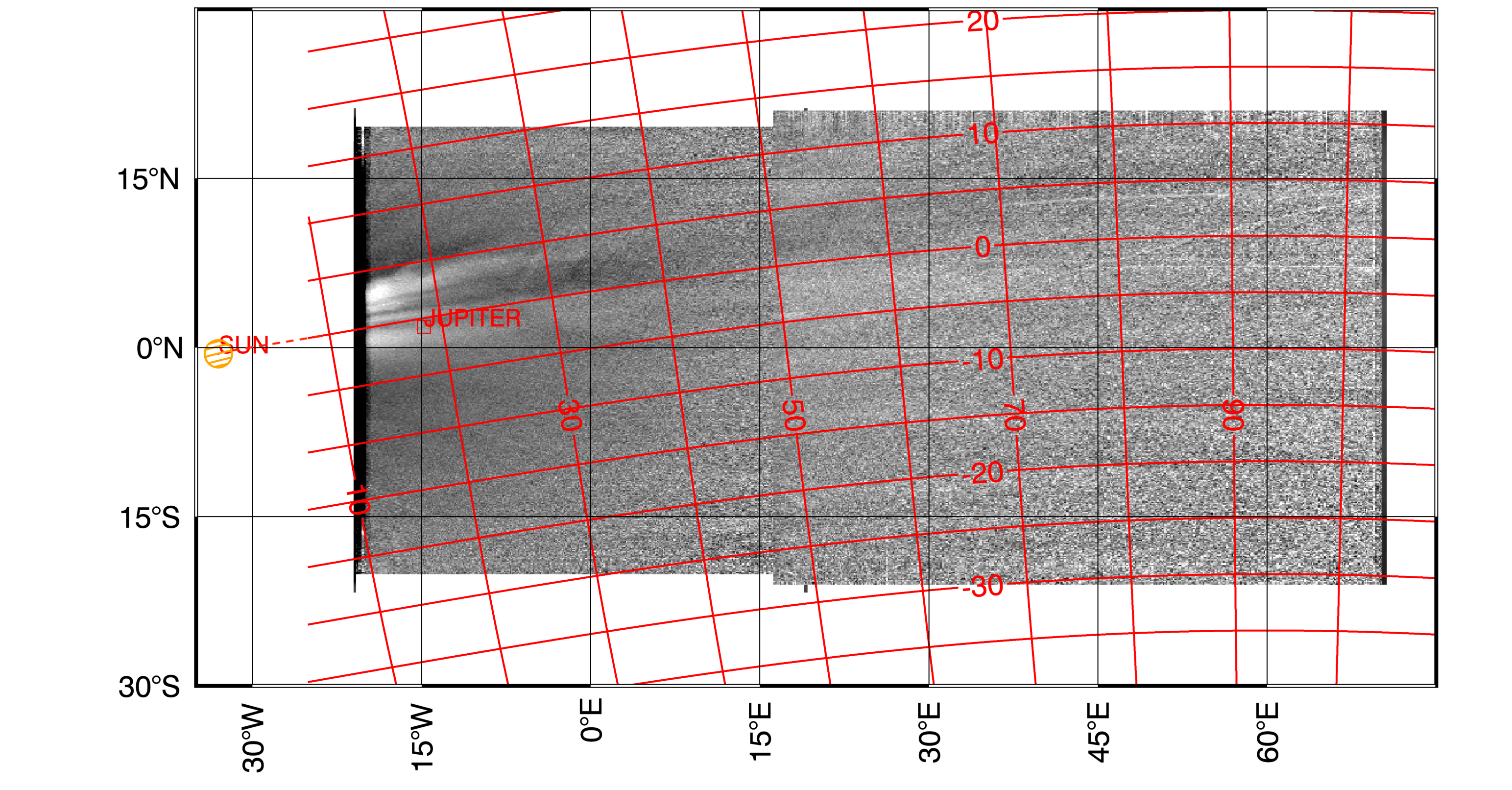}
\includegraphics[width=1.0\textwidth,clip=]{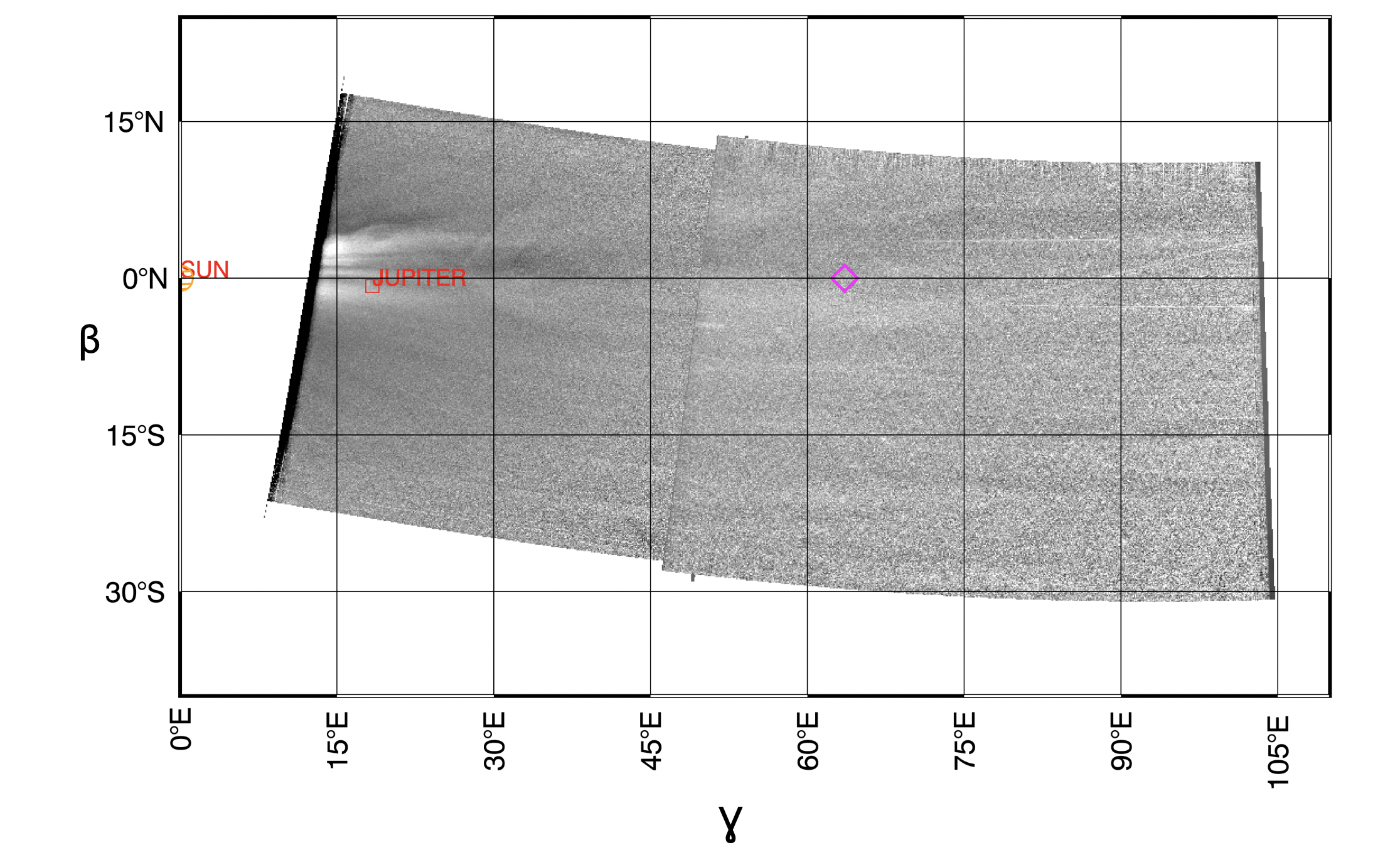}

              \caption{Top: Combined WISPR-I and WISPR-O images of a CME in the inner telescope
camera frame (black grid lines) in which the two angles, which define the lines-of-sight, are measured relative 
to the inner telescope centerline at (0,0). The Sun is shown 13$^{\circ}$ further west than the inner edge of the FOV.  
The red grid is an overlay of the $\gamma-\beta$  grid lines of the PSP orbital  frame where $\beta = 0$ is PSP's 
orbit plane,  $\gamma$ is the angle in the plane,  $\beta$ is the angle out of the plane, and the Sun is 
at  ($\gamma, \beta) = (0,0)$.  Bottom: The same images re-projected into the $\gamma-\beta$  
PSP orbital  frame (black grid lines). The magenta diamond marks the direction of the PSP velocity vector at this time  
(2018 November 1 at 22 UTC).}.
   \label{fig:6}
   \end{figure}

The small CME of 2019 April 1-2 was shown at four times in Figure~\ref{fig:5}, where the images are running-differenced L2 images from WISPR-I and
the tracked feature, the lower dark spot, is marked (red X).
A video of this event, made from the differenced images, is 
included in the online version.  We identified the probable source of this CME as AR 12737, which was seen to emerge on 2019 March 31 at 
approximately 13 UTC by both STEREO-A (STA hereafter)
and the {\it Solar Dynamics Observatory} (SDO). The region was officially identified as AR 12737 with HCI coordinates (longitude, latitude)  = (87.5$^{\circ}$, 12$^{\circ}$) at 2019 April 2 00 UTC, implying that at the time of emergence, the HCI 
longitude of the region was approximately 67$^{\circ}$. The flux rope was seen by PSP, STA and SOHO.

\begin{figure}    
   \centerline{\includegraphics[width=1.0\textwidth,clip=]{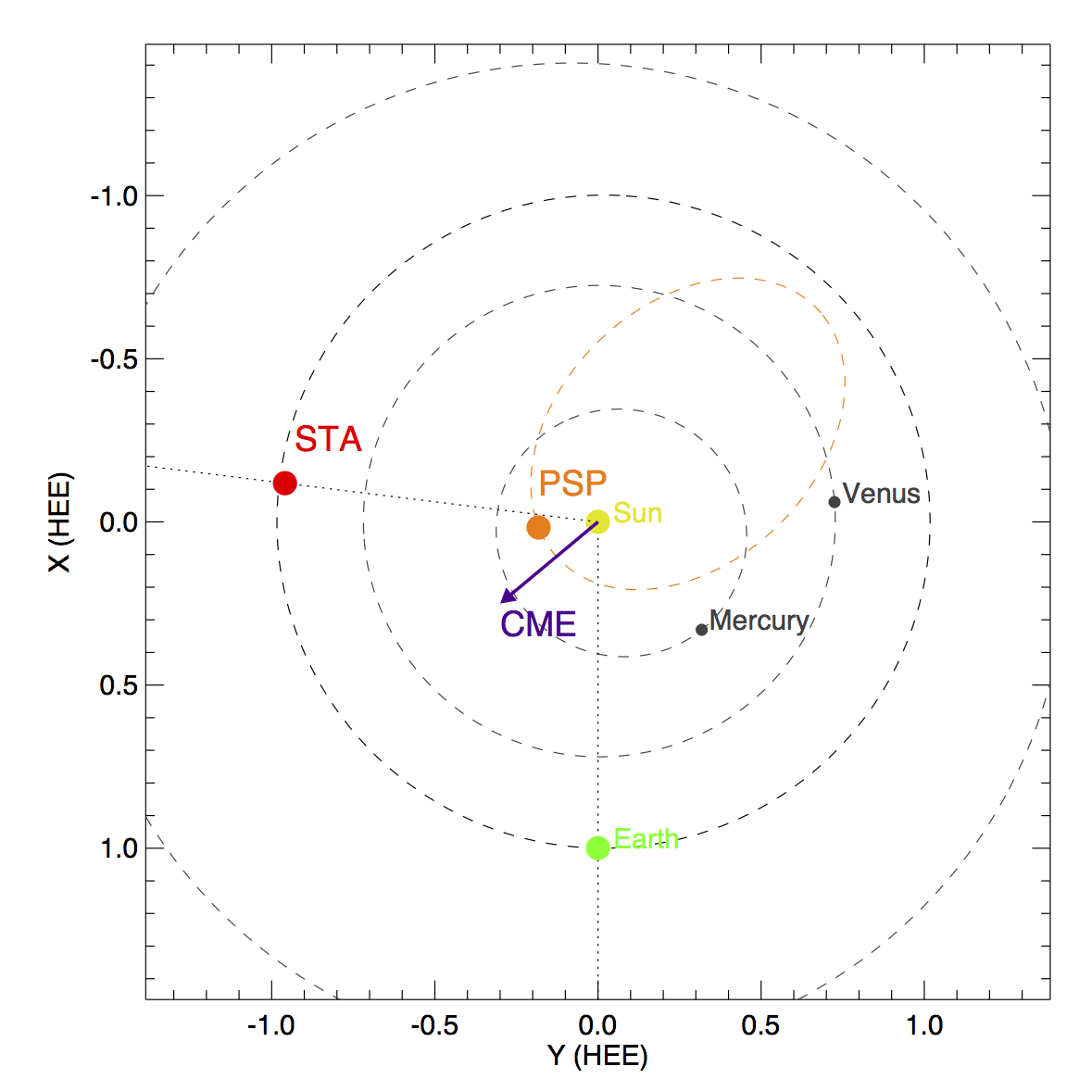}
              }
              \caption{Trajectory of 2019 April 2 flux rope in relation to PSP,
STEREO-A and Earth, projected in the Heliocentric Earth Ecliptic (HEE) reference
frame. The  tracking/fitting solution gave the CME propagation direction to be HCI longitude $= 67^{\circ} \pm 1^{\circ}$. 
Note that the purple arrow only indicates the direction of the CME, and is not meant to indicate the distance to the Sun. The HCI longitudes of the Earth and STA 
are 117$^{\circ}$ and 19$^{\circ}$, respectively. The orange dashed ellipse is PSP's orbit. }
   \label{fig:7}
   \end{figure}

A sequence of about 20 running-difference images, covering 12 $-$ 18 UTC on 2019 April 2, was used to track the flux rope, including the four in Figure~\ref{fig:5}. 
The flux rope's motion was out of the PSP orbit plane, and the full two-step fitting procedure, as described in Section 2.2.1, 
was used to solve for the trajectory. The results of the fitting procedure for this case were discussed in Section 2.2.1 and the 
plots shown in Figure~\ref{fig:ex1}. The resulting trajectory parameters were 
HCI longitude $= 67^{\circ} \pm 1^{\circ}$, 
HCI latitude $ = 6.0^{\circ} \pm 0.3^{\circ}$, $V = 333 \pm 1$ km s$^{-1}$, and $r_2(t_0) /R_{\odot} = 13.38 \pm 0.01$ with $t_0 = 12:09$~UTC on April 2. 
The error bars are determined from the uncertainty in the $\gammabeta$ inputs to the
fitting program as described in Section 2.2.1.  Additional sources of error from the tracking itself are discussed in Section 4.
Figure~\ref{fig:7} shows the trajectory vector in relation to locations of PSP, STA and Earth at 2019 April 2 at 18:09 UTC. Assuming 
the CME traveled radially (fixed HCI longitude), the HCI longitude determined for this flux rope ($67^{\circ} \pm 1^{\circ}$) is consistent 
with originating from AR12737 shortly after emergence since the region emerged at approximately HCI longitude $= 67^{\circ}$.

\begin{figure}    
\includegraphics[width=0.95\textwidth,clip=]{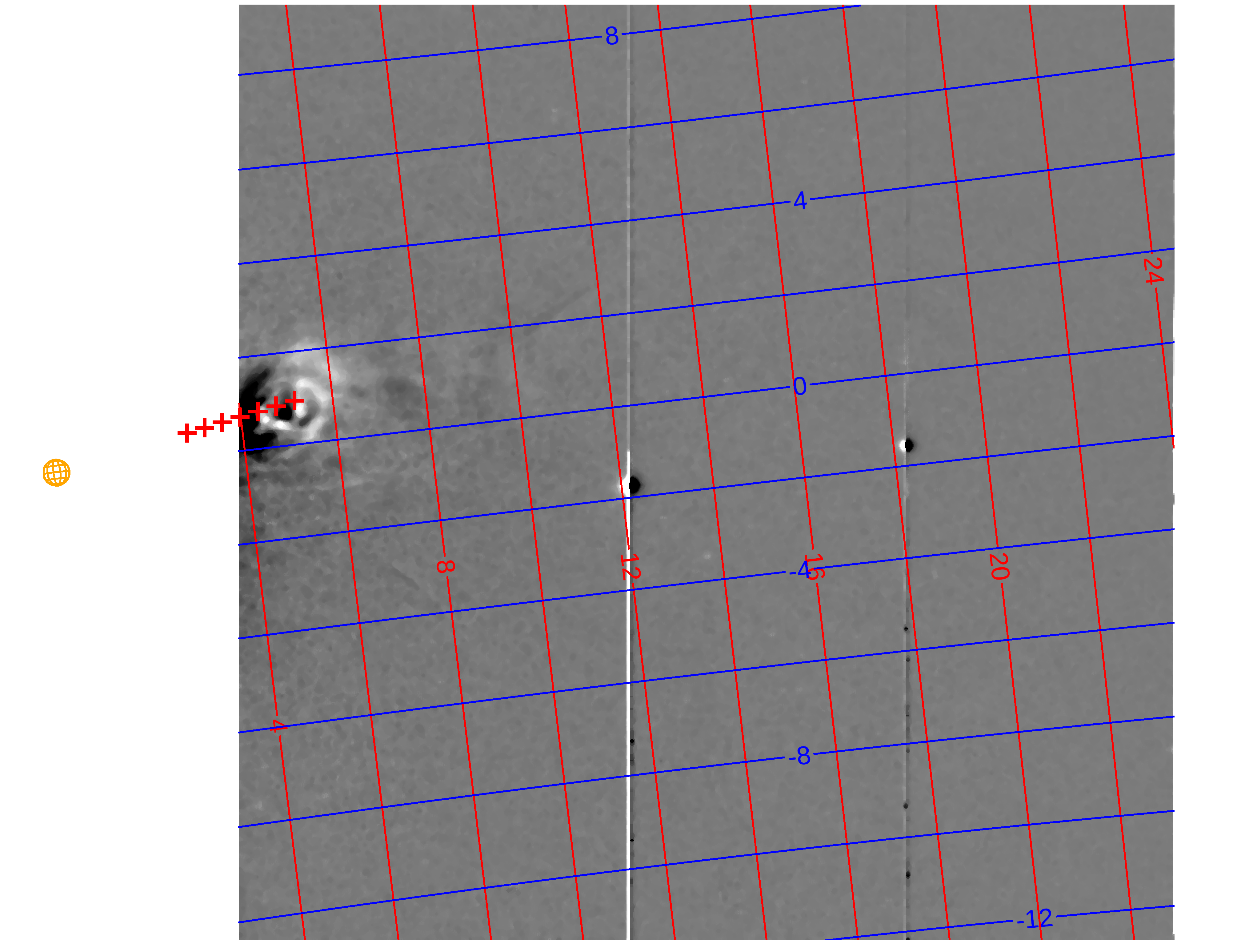}
\includegraphics[width=0.95\textwidth,clip=]{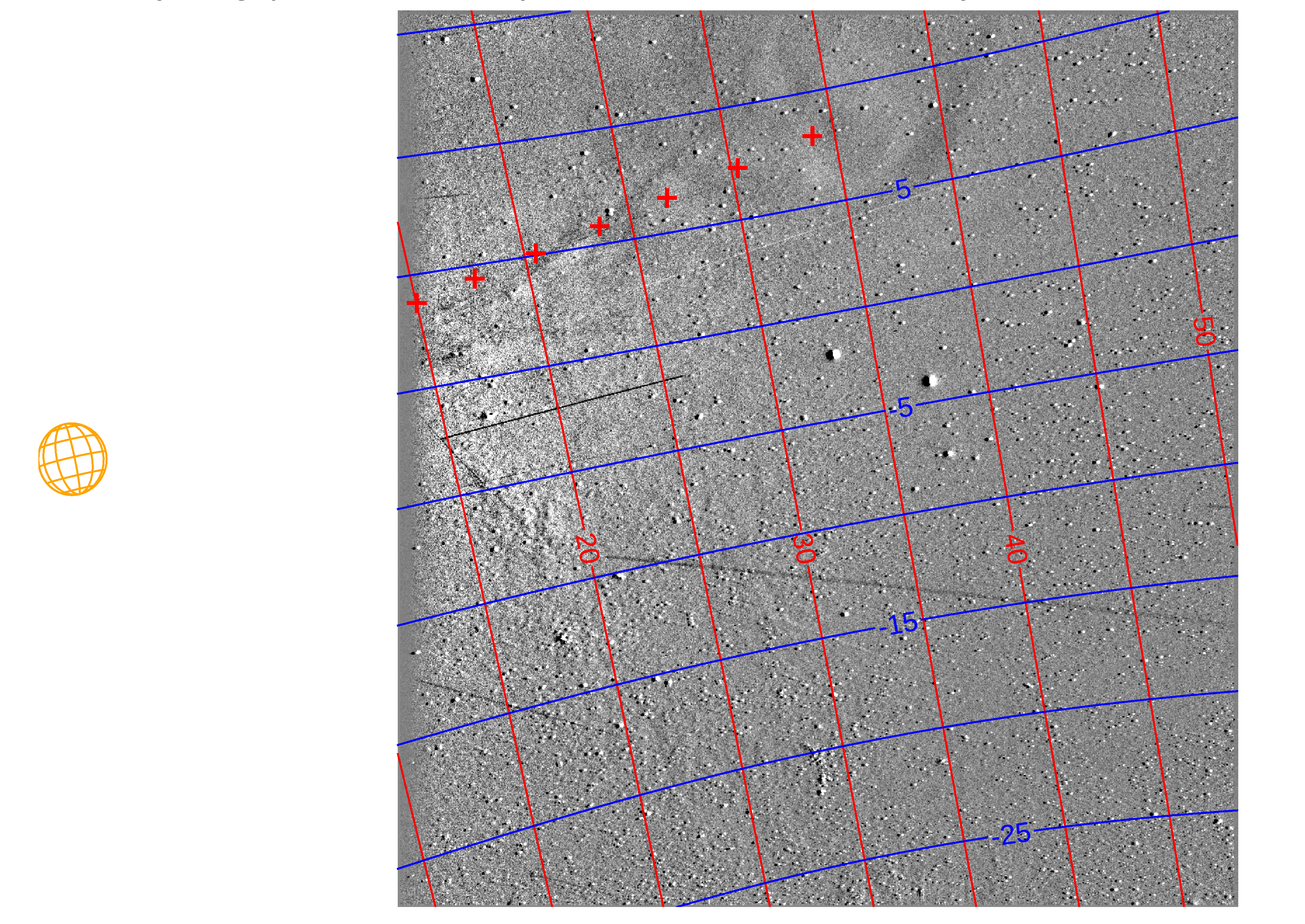}
              \caption{Trajectory of the flux rope on 2019 April 2 predicted using the tracking/fitting technique projected to images from
STEREO-A/HI-1 (top, 18:09 UTC), and WISPR-I (bottom, 18:13 UTC). The predicted trajectory (red +'s)  is shown from 12:09 to 
18:09 UTC in hourly increments. The location of the prediction for the last time (18:09 UTC) is in good agreement with the
location of the tracked feature seen in the HI-1A image, as well as the WISPR image. In both images, the grid lines show helioprojective coordinates.
In both images, the size and location of the Sun (yellow globe) are shown to scale.}
   \label{fig:8}
   \end{figure}

To validate the trajectory results from the tracking/fitting procedure, we make use of observations of this flux rope 
by a second white-light imager by predicting where the feature should appear in images 
from the other telescope. Here, we use STEREO-A/HI1 images because the CME structure was much better defined in these images than 
in those from COR2A. The predicted trajectory is computed in HCI coordinates at several times, then projected onto the image 
plane of the other telescope using the information in the image's FITS header (WCS, camera model, spacecraft location etc.). 
The top panel of Figure~\ref{fig:8} shows the predicted locations as red symbols (+) on 2019 April 2 from 12:09 to 18:09 UTC 
in hourly increments, projected onto an HI1A image at 18:09 UTC. The image time is close to the time of the last trajectory point and it  can be seen the 
predicted location of the tracked dark feature in the HI1A image is quite close to the observed location. This validates the results from the tracking/fitting procedure. The lower panel of Figure~\ref{fig:8} shows the same 
predicted trajectory locations projected back onto a nearly simultaneous WISPR image (18:13 UTC), one of those used in the tracking. 

Nearly simultaneous image pairs from WISPR and HI1A were also used to determine the location of the flux rope using triangulation to provide 
another test of the tracking/fitting technique. Triangulation makes no assumption about the trajectory of 
the feature; it is limited  by the ability to identify the same feature in both images. Performing triangulation at four 
times on April 2 from 16 to 18 UTC (including using the two images in Figure~\ref{fig:8}) gave the HCI longitude of the flux 
rope to be $67 \pm 1^{\circ}$, in excellent agreement with the longitude of $67 \pm 1^{\circ}$ found by tracking and 
fitting. Comparing the radii from triangulation to the predicted radii, the average discrepancy for the four times was 0.75 $R_{\odot}$.

\subsection{Flux rope of 2018 November 1-2}

\begin{figure}    
   \centerline{\includegraphics[width=1.0\textwidth,clip=]{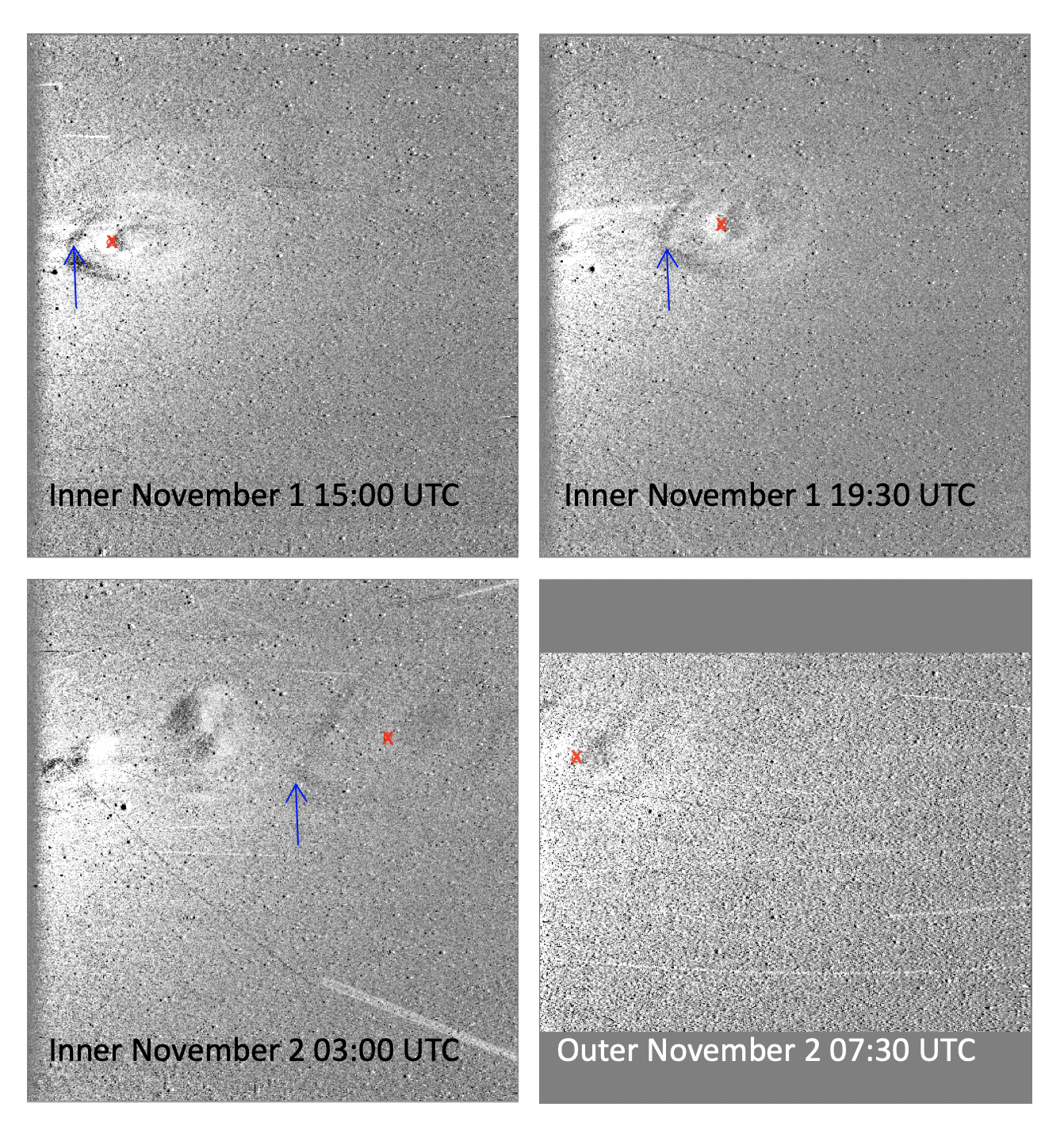}
              }
              \caption{WISPR running-difference images  at four
times for the CME of 2018 November 1-2; the first three are from WISPR-I and the last from
WISPR-O. The pixel location of first tracked feature (the back edge of the dark cavity) is marked with red X's. The blue
arrow points to a second feature tracked, the tail end of the flux rope, to compare with tracking/fitting results of
the first feature (see Section 4).}
   \label{fig:9}
   \end{figure}

\begin{figure}    
   \centerline{\includegraphics[width=1.0\textwidth,clip=]{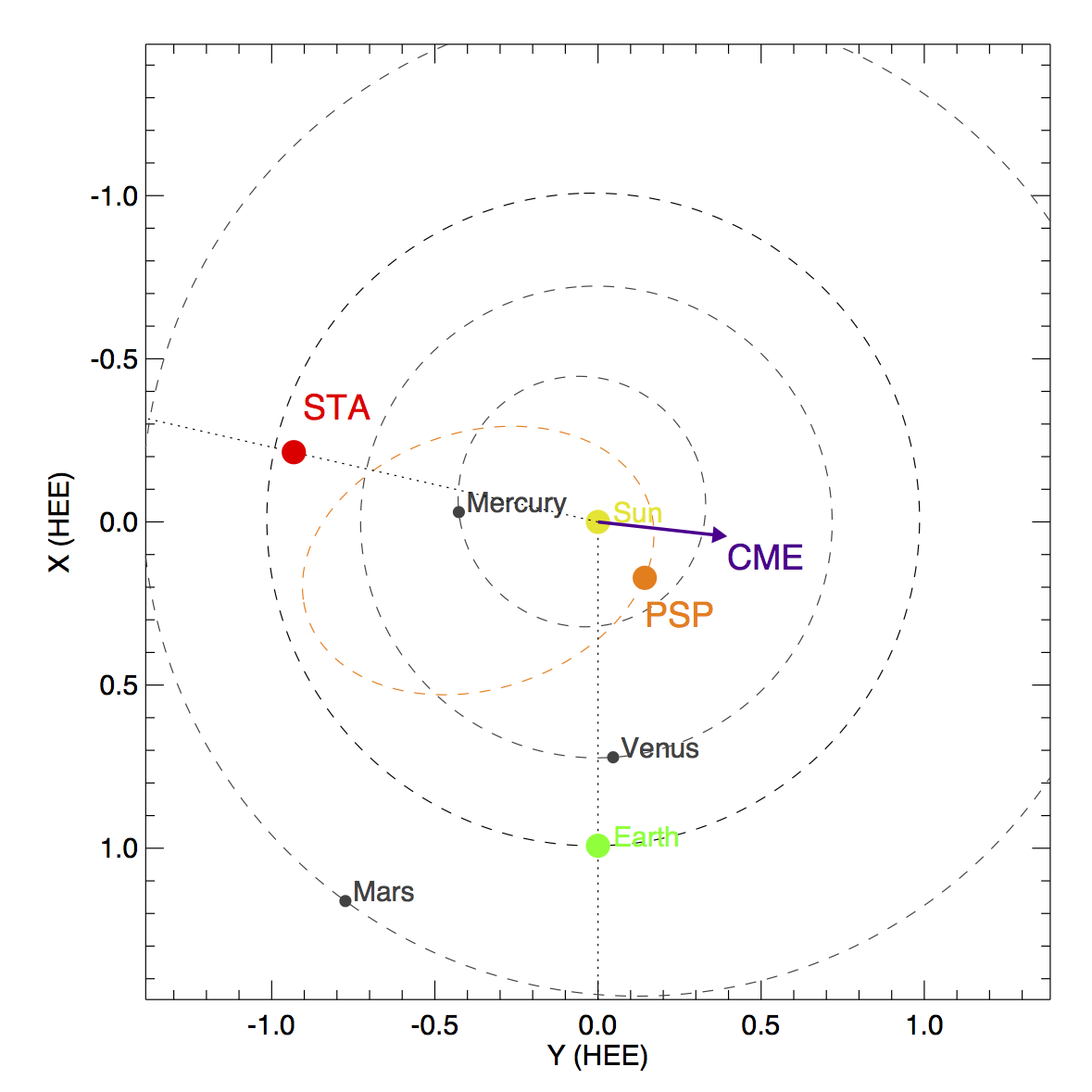}
              }
              \caption{Plot showing the direction of the flux rope, as determined
by the tracking/fitting technique, with respect to PSP, STEREO-A, and Earth for 2018 November 1 at
17:15 UTC, projected in the Heliocentric Earth Ecliptic (HEE) reference
frame. The CME propagation direction was found  to be HCI longitude $ = 47^{\circ} \pm 2^{\circ}$. Note that the purple arrow only indicates the direction of the CME, and is not meant to indicate the distance to the Sun.
The HCI longitudes of Earth and PSP are $-37^{\circ}$ and 3$^{\circ}$, respectively. The orange dashed ellipse is PSP's orbit.}
   \label{fig:10}
   \end{figure}

The CME of 2018 November 1-2 was the first flux rope observed by WISPR \citep{Howard2019}. It was analyzed in detail by \citet{Hess2020}; 
the online version of the later paper includes an animation of the flux rope. This flux rope was also 
observed by LASCO, with both observing it for about 12 hours on November 1. Figure~\ref{fig:9} 
shows four of the running-difference images used in tracking the flux rope with red X's marking the first feature tracked, the back of the dark flux rope cavity
on the line separating the cavity from the brighter compressed material following. (The blue arrow indicates a second feature tracked, discussed below in Section 4.) The first three 
images are from WISPR-I images and the last one from WISPR-O.   This feature was 
visible to WISPR for about 25 hours from November 1 at 12:47 UTC to November 2 at 17:17 UTC, crossing from the inner to outer telescope 
at about 06 UTC on November 2, and was tracked in about two-dozen images.  This feature was somewhat difficult to track in 
the outer telescope image, as evidenced by the fourth image in Figure~\ref{fig:9}. For this CME, motion was very close to the 
PSP orbit and solar equatorial planes and, as discussed in Section 2.2.2, the two-step fitting procedure defaults to a 
one-step fitting procedure.  The results of the fitting program were shown in Figure~\ref{fig:ex2}.  
The trajectory solution from the fitting program was $V = 218 \pm 2$ km s$^{-1}$, HCI longitude $ = 47^{\circ} \pm 2^{\circ}$; HCI latitude 
$=  -1^{\circ} \pm 2^{\circ},  r_2(t_0)/R_{\odot} =  15.7 \pm 0.2$ where $t_0$ is November 1 at 12:47 UTC, the 
time of the first image used in the tracking. The error bars are determined from the uncertainty in the $\gammabeta$ inputs to the
fitting program as described in Section 2.2.2.  
Figure~\ref{fig:10} shows the vector of this trajectory in relation to the locations of Earth, STA, and 
PSP on November 1 at 17:15  UTC. Earth was at HCI longitude $= -37^{\circ}$. Thus the flux rope was 84$^{\circ}$ from Earth, 
consistent with the observation that this CME was a limb event as seen from Earth \citep{Hess2020}.

\begin{figure}    
   \includegraphics[width=0.95\textwidth,clip=]{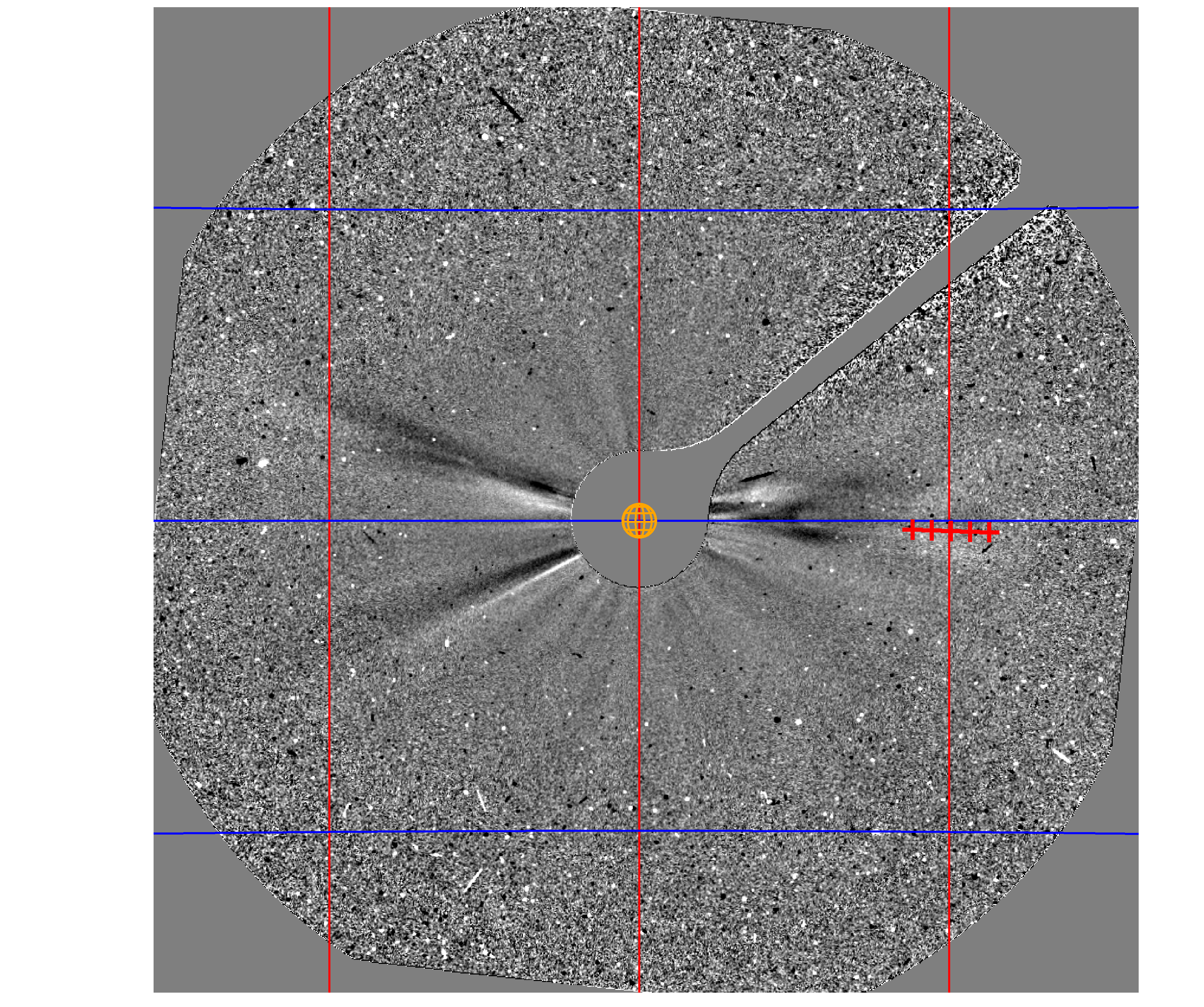}
   \includegraphics[width=0.95\textwidth,clip=]{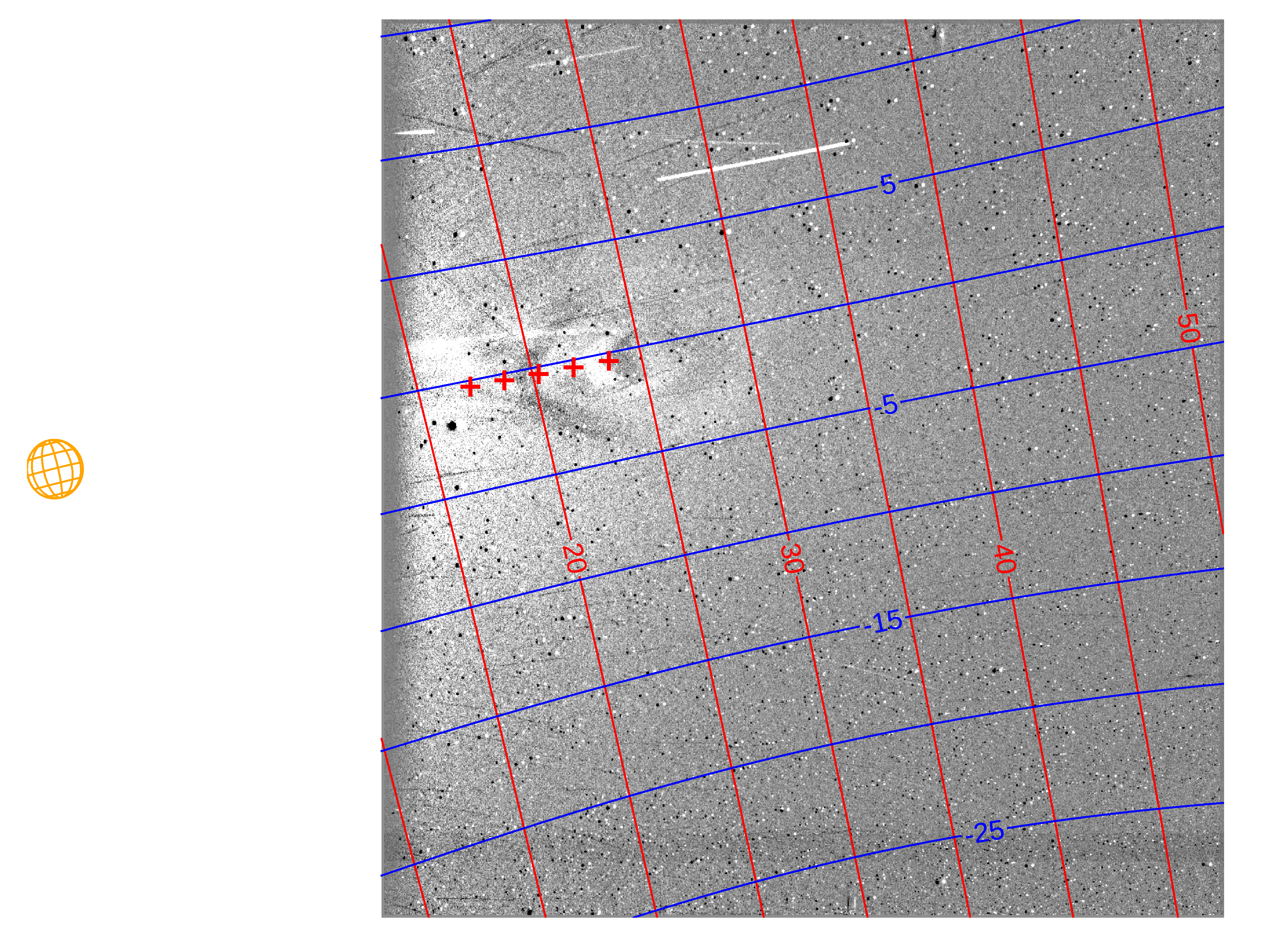}
              \caption{Top: Predicted location of the back of the flux rope cavity for 2018 November 1 from 13:15 to 
17:15 UTC in hourly increments projected onto a LASCO/C3 image at 17:16 UTC. The predicted distance of the feature from Sun for C3 is off by about 2 hrs or 2 R$_{\odot}$. Bottom: the same predicted trajectory points are plotted on a WISPR-I image at 17:15 UTC, used in 
tracking the flux rope. In both images, the grid lines show helioprojective coordinates.}
   \label{fig:11}
   \end{figure}

As for the flux rope in Section 3.1, we compare the location as predicted from the trajectory solution to what was seen from another white-light imager 
with a second near-Earth viewpoint, SOHO/LASCO C3. The top panel of Figure~\ref{fig:11} shows the prediction from tracking 
the rear of the flux rope cavity (the first feature) projected onto a C3 coronagraph image from 2018 November 1 at 17:16 UTC. The HCI 
coordinates of this feature as predicted by the trajectory solution were computed hourly starting on November 1 at 13:15 UTC 
and ending at 17:15 UTC, the time of the images.  While the prediction is reasonably close to the trajectory as observed by C3, the 
feature can be seen to lag the prediction by about 2 hours. This corresponds to approximately  2 R$_{\odot}$ at a time when the 
predicted distance for the Sun was 20 R$_{\odot}$. This discrepancy is discussed in Section 4. 
The right panel shows the same predicted trajectory points plotted on a nearly simultaneous WISPR image from 17:15 UTC, one of the images used in
tracking.

As another test of the tracking/fitting method, triangulation was performed using four nearly simultaneous C3 and WISPR-I image pairs, 
covering seven hours, including the pair in Figure~\ref{fig:11}. Triangulation between WISPR and C3 images requires no knowledge of 
the trajectory, but does assume that the  same ``feature" can be identified  in both images. The HCI longitude was determined to be
$59^{\circ} \pm 6^{\circ}$ compared to the fitting/tracking 
solutions of HCI longitude $47^{\circ} \pm 2^{\circ}$. The predicted distance from the Sun at 17:15 UTC, the time of the images in Figure~\ref{fig:11}, 
was $r/R_{\odot}= 19  \pm 1$ compared to the predicted distance of $r/R_{\odot}= 21  \pm 1$. Thus the distances agree within the uncertainties, 
but the longitudes do not.  The possible sources of the discrepancy will be further discussed in the next section.

\section{Summary and Discussion}

In this paper, we have presented a tracking/fitting technique for determining the trajectory of coronal ejecta, assumed 
to be moving radially at a constant velocity, that are observed in a sequence of images by WISPR on {\it Parker Solar Probe}.  
Although WISPR has a fixed angular field-of-view, the physical extent of the coronal region imaged changes with time 
due to PSP's highly elliptical orbit. Thus an object's observed motion in a sequence of images is a combination of its 
intrinsic motion and the spacecraft motion, making new analysis techniques necessary. We presented the 
two equations relating an object's position in an image to its coordinates in a heliocentric inertial frame that 
are valid for observations from a rapidly moving spacecraft. The equations included a first-order correction for 
the inclination of the spacecraft orbit plane to the solar equatorial plane. Once the object is tracked in a 
sequence of images, a procedure that fits the track to these two equations is used to determine the 3D trajectory: 
distance vs. time, longitude, and latitude. The fitting procedure is somewhat different for objects moving in or out of the orbit plane.

Results from tracking two flux ropes seen by WISPR were presented. Both flux ropes were observed by another white-light imager, 
and the fitting/tracking technique was validated and its uncertainty investigated using observation of these 
events from the second viewpoint.

The first was a small CME seen during the second solar encounter on 2019 April 2 whose motion was out of the PSP orbit plane.  
Several tracks  were made and the final result uses the variations to determine the uncertainties. The results 
for the trajectory parameters were 
HCI longitude $= 67^{\circ} \pm 1^{\circ}$, 
HCI latitude $ = 6.0^{\circ} \pm 0.3^{\circ}$, $V = 333 \pm 1$ km s$^{-1}$, and $r_2(t_0) /R_{\odot} = 13.38 \pm 0.01$ with $t_0 = 12:09$~UTC on April 2.
This CME was also observed by STEREO-A. Using four sets of 
nearly simultaneous observations from WISPR and HI1A, the position of the object was determined using triangulation, 
which makes no assumptions about the trajectory. The solution for the HCI longitude and distance were in excellent agreement 
with the result from tracking and fitting for these times within the  uncertainties. 

The second, larger flux rope was the first CME seen by WISPR on 2018 November 1-2; its motion was very close to the orbit plane. 
From tracking the back of the CME cavity we obtained the trajectory parameters 
 $V = 218 \pm 2$ km s$^{-1}$, HCI longitude $ = 47^{\circ} \pm 2^{\circ}$; HCI latitude 
$=  -1^{\circ} \pm 2^{\circ},  r_2(t_0)/R_{\odot} =  15.7 \pm 0.2$ where $t_0$ is November 1 at 12:47 UTC, the 
time of the first image used in the tracking. The errors are based on the uncertainty in the input data to the tracking/fitting solution. 
This CME was observed by SOHO/LASCO; as for 
the first CME, we used nearly simultaneous images from the two viewpoints to find position by triangulation. Comparing 
the results and uncertainties from the tracking/fitting technique with the results from triangulation, we found discrepancies larger 
than the error bars from the tracking/fitting solution.  

For the second flux rope, the difference between predicted and observed locations seen in the C3 comparison in 
Figure~\ref{fig:11} and the difference between predicted and triangulated longitudes are larger than the error bars 
in the fitting/tracking solution as determined by uncertainties in the $\gammabeta$  inputs (see Section 2). 
These solution uncertainties were found to be very small ($\pm 2^{\circ};  \pm 1 R_{\odot}$), and are much smaller than typical errors found in trajectory 
 determination of CMEs tracked in STEREO/SECCHI data \citep[cf,][]{Lugaz2010, Liewer2011}. 
The primary cause of errors for trajectory determination by 
tracking diffuse CMEs is line-of-sight effects \citep[e.g.,][]{Lugaz2010}, 
which limit the ability to identify and track the same ``feature" in white light images from different 
times or viewpoints. For  this flux rope, PSP and Earth were separated by $40^{\circ}$, whereas for the first flux rope PSP and STA were nearly aligned (cf. Figure 7). Tracking assumes the same point in 3D space as been identified in the images, yet the features seen in white light are 
the result of integration of the Thomson-scattering signal from all electrons along the of line-of-sight. The same CME feature may look 
different from different viewpoints or at different times as the structure is seen from different angles or as it evolves. On the other hand, the equations that form the basis of the tracking/fitting solution were derived for the motion of a single point. 
 
To better understand how much such effects might influence the reliability of the tracking/fitting solutions, we tracked a second feature of the 2018
November 1-2 CME, the tail end of the flux rope, identified in Figure~\ref{fig:9} images by the blue arrow.
For this feature, the trajectory solution from the tracking/fitting program 
was $V = 190 \pm 6$ km s$^{-1}$, HCI longitude $= 52^{\circ} \pm 1^{\circ}$; HCI latitude $= -2^{\circ} \pm 3^{\circ}, r_2(t_0)/R_{\odot} 
=  13 \pm 1$ where $t_0$ is November 1 at 14:17 UTC, the time of the first image used in the tracking. This suggests that the uncertainty 
in the CME longitude of this event could be as large as $\pm 5^{\circ}$. 
For the 2019 April 1-2 CME, we also applied this same practice to track a second feature near the primary feature studied in Section 3.1, 
and found the second feature's longitude different from the primary feature by 4$^{\circ}$. 
Such exercise provides another estimate of the uncertainties in the tracking/fitting technique, although
it should be noted that different features in the same CME may exhibit different motion patterns due to evolution of the CME structure.

The difficulty in tracking the same feature in a series of images of a diffuse CME is a limitation on the reliability of the tracking/fitting
technique and its validation. 
The assumption of radial propagation at constant velocity is also a 
serious limitation of the technique. CMEs certainly accelerate, 
but most of the acceleration occurs below 2~$R_{\odot}$ \citep{Patsourakos2010, Zhu2020}, lower than WISPR observes. Observations 
of strong transverse flows by the {\it in situ} instruments on PSP \citep[e.g., ][]{Howard2019} have suggested that the corona co-rotates 
with the Sun out to the distances observed by PSP.  Density features embedded in streamers that are attached to 
the Sun then may be effected, violating the assumption of radial motion. Our technique can also be extended to assume constant 
radial motion in the Carrington coordinate frame and this will be investigated further in the future.

\begin{acks}
We gratefully acknowledge the help and support of William Thompson, Adnet System, Inc. and the WISPR team throughout this work. We also thank Guisseppe Nistico, Marco Velli and Brian Wood for helpful conversations. Parker Solar Probe was designed, built, and is now operated by the Johns Hopkins Applied Physics Laboratory as part of NASA's Living with a Star (LWS) program (contract NNN06AA01C). 
We thank the reviewer for valuable comments which have improved the clarity of the presentation.
The work of PCL, JRH and PP was conducted at the Jet Propulsion Laboratory, California Institute of Technology under a contract from NASA. 
J. Q. is partly
supported by NASA's HGI program (80NSSC18K0622). A.V. is supported by the WISPR Phase E program at APL. RAH  is supported by a contract from the NASA PSP program for the WISPR program.  The WISPR L2  FITS images used in this study are available at \verb!https://wispr.nrl.navy.mil/wisprdata!.

\end{acks}

\appendix
\section{Transformation of Coordinates}
\label{A1} 

We transform a feature's position in a heliocentric frame (the $xyz$ frame), 
in which the $x-y$ plane is the solar equatorial plane, to its position 
in a heliocentric frame (the $x'y'z'$ frame), in which the $x'-y'$ plane is 
PSP's orbit plane and $z'$-axis is the direction of the orbital angular momentum
(Figure~\ref{fig:tran1}). For the transformation, we define the $x$-axis in the $xyz$ frame to be the 
ascending node of PSP's orbit relative to the solar equatorial plane, and the $x'$-axis 
pointing from the Sun to PSP. The inclination of PSP's orbit $\epsilon$ is 
the amount of rotation of PSP's orbital plane about the $x$-axis. The spacecraft's 
motion in its own orbital plane can be described by its distance to the Sun $r_1$ 
and the angle $\phi_1$ measured from the $x$-axis. In other words, $\phi_1$ is the 
amount of rotation about the $z'$-axis. The feature's position in the $xyz$ frame 
is noted as $[x_2, y_2, z_2] \equiv [r_2\cos\delta_2\cos\phi_2, 
r_2\cos\delta_2\sin\phi_2, r_2\sin\delta_2]$, where $r_2$ is the distance to the Sun, $\phi_2$ is the 
angle in the $xy$ plane, measured from the $x$-axis, or the ascending node of PSP's orbit, and $\delta_2$ is
the angle with the $x-y$ plane. Note that $\phi_2$ in this frame is offset from the HCI longitude by
a constant, which is the HCI longitude of the ascending node of PSP's orbit relative to the solar equatorial
plane, and this constant is determined from the ephemeris. The feature's position in the $x'y'z'$ frame is noted 
as $[x'_2, y'_2, z'_2] \equiv [r_2\cos\delta'_2\cos\phi'_2,
r_2\cos\delta'_2\sin\phi'_2, r_2\sin\delta'_2]$ (see Figure~\ref{fig:tran1}), 
and the transformation from $(x_2, y_2, z_2)$ to $(x'_2, y'_2, z'_2)$ is given by
\begin{eqnarray}
& \left[ \begin{array}{c} x'_2 \\ y'_2 \\z'_2 \end{array} \right]
= \begin{bmatrix} \cos\phi_1 & \sin\phi_1 & 0 \\ -\sin\phi_1 & \cos\phi_1 & 0 \\ 0 & 0 & 1 \end{bmatrix}
\begin{bmatrix} 1 & 0 & 0 \\ 0 & \cos\epsilon & \sin\epsilon \\ 0 & -\sin\epsilon & \cos\epsilon \end{bmatrix}
\left[ \begin{array}{c} x_2 \\ y_2 \\z_2 \end{array} \right] & \nonumber \\
& = \left[ \begin{array}{c} r_2(\cos\delta_2\cos\phi_2\cos\phi_1 + \cos\delta_2\sin\phi_2\sin\phi_1\cos\epsilon + \sin\delta_2\sin\phi_1\sin\epsilon) \\
r_2(-\cos\delta_2\cos\phi_2\sin\phi_1 + \cos\delta_2\sin\phi_2\cos\phi_1\cos\epsilon + \sin\delta_2\cos\phi_1\sin\epsilon)  \\
r_2(-\cos\delta_2\sin\phi_2\sin\epsilon + \sin\delta_2\cos\epsilon) \end{array} \right] .
\end{eqnarray}

We can find the relationship between the $\gamma-\beta$ coordinates and the feature's position $[r_2, \phi_2, \delta_2]$
in the $xyz$ frame,
\begin{flalign}
 \frac{\tan\beta}{\sin\gamma} & = \frac{z'_2}{y'_2} & \nonumber \\
 & = \frac{-\cos\delta_2\sin\phi_2\sin\epsilon + 
\sin\delta_2\cos\epsilon}{-\cos\delta_2\cos\phi_2\sin\phi_1 + \cos\delta_2\sin\phi_2\cos\phi_1\cos\epsilon + \sin\delta_2\cos\phi_1\sin\epsilon}, & \label{eq:a1}
\end{flalign}
and
\begin{flalign}
 \cot \gamma & = \frac{r_1 - x'_2}{y'_2}&  \nonumber \\
 & = \frac{r_1 - r_2(\cos\delta_2\cos\phi_2\cos\phi_1 + \cos\delta_2\sin\phi_2\sin\phi_1\cos\epsilon + 
\sin\delta_2\sin\phi_1\sin\epsilon)}{r_2(-\cos\delta_2\cos\phi_2\sin\phi_1 + \cos\delta_2\sin\phi_2\cos\phi_1\cos\epsilon + 
\sin\delta_2\cos\phi_1\sin\epsilon)}. & \label{eq:a2}
\end{flalign}
In the above equations, $\gamma$, $\beta$, $\phi_1$, $r_1$, and $r_2$ are time dependent, and $\phi_2, \delta_2$, and $\epsilon$ are constant. For convenience, 
we ommit the $t$ dependence in the expressions of relevant properties.
For a small angle $\epsilon$, we expand Equations~\ref{eq:a1} and \ref{eq:a2} and keep the first-order terms of $\epsilon$ to arrive at
\begin{flalign}
& \frac{\tan\beta}{\sin\gamma} = \frac{\tan\delta}{\sin(\phi - \phi_1)} \left(1 - F\sin\epsilon\right), & \label{eq:a3}
\end{flalign}
(same as the Equation~\ref{eq:3} in Section 2.1), and 
\begin{flalign}
& \cot \gamma = \frac{r_1- r_2\cos\delta_2\cos(\phi_2-\phi_1)}{r_2\cos\delta_2\sin(\phi_2 - \phi_1)}
\left(1 - G\sin\delta_2 \sin\epsilon\right), & \label{eq:a4}
\end{flalign}
(same as Equation~\ref{eq:4} in Section 2.1) where the coefficients of the first-order terms are given by 
\begin{flalign}
& F(\phi_2,\delta_2, \phi_1) \equiv \frac{\sin\phi_2}{\tan\delta_2} + \frac{\tan\delta_2\cos\phi_1}{\sin(\phi_2 - \phi_1)}, & \label{eq:F}
\end{flalign}
and
\begin{flalign}
& G(r_2, \phi_2, \delta_2, r_2, \phi_1) \equiv \frac{\sin\phi_1}{r_1/r_2 - \cos\delta_2 \cos(\phi_2-\phi_1)} + \frac{\cos\phi_1}{\cos\delta_2 \sin(\phi_2 - \phi_1)} &\label{eq:G} &
\end{flalign}

In addition, the angle $\beta$ can be also independently derived from trigonometry,
\begin{flalign}
\tan \beta & = \frac{z'_2}{\sqrt{x^{'2}_2+y^{'2}_2 + r_1^2 - 2x'_2r_1}}  & \nonumber \\
& = \frac{r_2\sin\delta_2}{\sqrt{r_2^2\cos^2\delta_2 + r_1^2 - 2r_2r_1\cos\delta_2\cos(\phi_2 - \phi_1)}} 
 \left( 1 - H\sin\epsilon\right), & \label{eq:a5}
\end{flalign}
where the coefficient for the first-order term is 
\begin{flalign}
& H(\phi_2,\delta_2, r_2, \phi_1, r_1) \equiv \frac{\cos\delta_2\sin\phi_2}{\sin\delta_2} + \frac{r_2^2\cos\delta_2\sin\delta_2\sin\phi_2 - 
r_2r_1\sin\delta_2\sin\phi_1}{r_2^2\cos^2\delta_2 + r_1^2 - 2r_2r_1\cos\delta_2\cos(\phi_2 - \phi_1)}. & \label{eq:H}
\end{flalign}

A series of $\gammabeta$ measurements in the PSP orbital frame at times $t_i$ 
are fit to Equations~\ref{eq:a3} and \ref{eq:a4} to determine parameters characterizing 
particle's motion in the HCI frame. Equation~\ref{eq:a5} is used to calculate $\beta(t_i)$ from
fitting parameters and compare with observed $\beta(t_i)$.

\section{Initial Guess of Fitting Parameters}
\label{A2} 

The convergence of a non-linear least squares fit often depends on the initial input.
For out-of-plane motions, we can calculate the
zeroth-order solutions of a feature's position $[r_2, \phi_2, \delta_2]$, 
and use them as the initial input for the fit. We first compute the initial 
guess of $\phi_2$ and $\delta_2$ using Equation~\ref{eq:1}, without considering the first-order corrections
of the orbit inclination. For this purpose, we define
\begin{flalign}
& \eta(t) \equiv \frac{\cos\phi_1(t)\tan\beta(t)}{\sin\gamma(t)} = \frac{\tan\delta_2}{\sin\phi_2-\cos\phi_2\tan\phi_1(t)}. & \label{eq:eta}
\end{flalign}
Taking measurements of $[\gamma(t), \beta(t)]$ at two times $t_i$ and $t_j$, we find
\begin{flalign}
& \phi_2 = \tan^{-1}\left[\frac{\eta(t_i)\tan\phi_1(t_i) - \eta(t_j)\tan\phi_1(t_j)}{\eta(t_i) - \eta(t_j) }\right], & \label{eq:phi}
\end{flalign}
and
\begin{flalign}
& \delta_2 = \tan^{-1}[\eta(t_i)(\sin\phi_2 - \cos\phi_2\tan\phi_1(t_i))]. & \label{eq:delta}
\end{flalign}

We note that, PSP usually only moves by a small angular distance during the observation; therefore,
$\eta(t)$ values are very close to each other, and the in-plane angle calculated using Equation~\ref{eq:phi} with only two images
is likely dominated by uncertainties in $\gamma(t)$ and $\beta(t)$ measurements.
To overcome this difficulty, we employ all $\gamma(t)$ and $\beta(t)$ measurements to compute the initial guess of $\phi_2$ with
a linear approximation. For a series of $\gammabeta$ measurements at times $t_i$, we 
define $\Delta \phi_1(t_i) \equiv \phi_1(t_i) - \bar\phi_1$, where $\bar\phi_1 \equiv \phi_1(t_m)$
is PSP's angle at a reference time $t_m$ when $\phi_1$ is closest to the median of $\phi_1(t_i)$ during the observation. For small
$\Delta \phi_1$, Equation~\ref{eq:phi} can be linearized,
\begin{flalign}
& 1 - \frac{\bar\eta}{\eta(t_i)} \approx \left(\frac{\sec^2\bar\phi_1}{\tan\phi_2 - \tan\bar\phi_1}\right) \Delta\phi_1(t_i) ,& \label{eq:linear}
\end{flalign}
where $\bar\eta$ is computed by Equation~\ref{eq:eta} at the reference time $t_m$. A least squares fit to the above linear relation
returns the scaling constant in the bracket, from which $\phi_2$ is computed and referred as $\bar\phi_2$.
The other angle $\delta_2$ is then computed applying Equation~\ref{eq:delta} to $\gamma - \beta$ measurements
at time $t_m$.

To estimate the range of $\phi_2$ values, we apply Equation~\ref{eq:linear} to a pair of $\gamma$ and $\beta$ measurements
obtained at the reference time $t_m$ and at any other time during the observation, which returns $N-1$ estimates of $\phi_2$, $N$ being
the total number of $\gamma$ and $\beta$ measurements. The standard deviation of these $\phi_2$ values, referred as
$\Delta \phi_2$, gives the range of the initial input $\phi_2$. As an example, for a flux rope observed by PSP on 2019 April 2 (see
Section 3.1 for the details of this event), $\bar\phi_2$ is found to be 75$^{\circ}$ (in HCI coordinates), and the
deviation $\Delta \phi_2$ is 42$^{\circ}$. These initial guesses are used in the first-step fit to Equation~\ref{eq:a3} to
determine the accurate values of $\phi_2$ and $\delta_2$ with the first order correction of the inclination of the orbit.

Subsequently, we use Equation~\ref{eq:2} to solve for $r_2(t)$,
\begin{flalign}
& r_2(t) = \frac{r_1(t)}{\cos\delta_2\left[\sin(\phi_2 - \phi_1(t))/\tan\gamma(t) + \cos(\phi_2 - \phi_1(t))\right]}. & \label{eq:r2}
\end{flalign}
With measurements of $\gamma$ and $\beta$ at two times $t_i$ and $t_j$, here
chosen to be the start and end times of the observation, we compute $r_2(t)$, and
solve for $r_{20}$ and $V$ to zeroth-order, 
\begin{flalign}
& V = \frac{r_2(t_j) - r_2(t_i)}{t_j - t_i}, & \\
&  r_{20} = r_2(t_i) - V (t_i - t_0). & \label{eq:vr0}  
\end{flalign}
These are used as initial input in the second-step fit to Equation~\ref{eq:a4} to determine $r_{20}$ and $V$ more accurately.

We conduct the Levenberg-Marquardt least-squares fit to Equations~\ref{eq:a3} and ~\ref{eq:a4} multiple times,
each time varying the initial guess of $\phi_2$ by 5$^{\circ}$ in the range from $\bar\phi_2 - \Delta \phi_2$ to $\bar\phi_2 + \Delta \phi_2$.
For each initial guess of $\phi_2$, the initial guesses of $\delta_2, r_{20}$, and $V$ are then derived, as described above.
Constraining the initial guess using the zeroth-order estimates helps the fit to converge.
In each fit, $\phi_2$ is allowed to vary from $-10^{\circ}$  to $+90^{\circ}$ around its initial guess, the range
of $\delta_2$ is $\pm 30^{\rm o}$ around the initial guess. In the second-step fit, in which $\phi_2$ and $\delta_2$
are fixed at the values determined from the first-step fit, the range of $r_{20}$ is between 1 and 35 solar radii, 
and the range of $V$ is between 10 to 2500 kilometers per second. The uncertainties of the final fitting results 
primarily depend on the uncertainties in the $\gammabeta$ measurements (see Section 2.2.1).

For in-plane motions when $\beta \sim 0$, Equation~\ref{eq:3} becomes trivial and cannot be applied to compute $\phi_2$.
Instead, $\gamma(t_i)$ measurements are fit to Equation~\ref{eq:4} to determine $\phi_2, r_{20}$, and $V$ altogether.
We conduct the one-step fit multiple times using 10 different initial guesses of $\phi_2$ spanning 180 degrees
starting from PSP's position $\phi_1$. In the fit, $\delta_2$ is fixed, and the fit is conducted multiple times
with $\delta_2$ varying from $-4^{\circ}$ to $4^{\circ}$ in 1$^{\circ}$ increments (see Section 2.2.2).

\newpage

\bibliographystyle{spr-mp-sola}
\bibliography{psp}


\newpage


\end{article}

\end{document}